\documentclass[fleqn,usenatbib]{mnras}

\usepackage[T1]{fontenc}
\usepackage{ae,aecompl}
\usepackage{xcolor}

\DeclareRobustCommand{\VAN}[3]{#2}
\let\VANthebibliography\thebibliography
\def\thebibliography{\DeclareRobustCommand{\VAN}[3]{##3}\VANthebibliography}

\usepackage{macros}
\usepackage{verbatim}

\usepackage{graphicx}	
\usepackage{amsmath}	
\usepackage{amssymb}	
\usepackage{booktabs}
\usepackage{array}
\usepackage{makecell}
\usepackage{ulem}
\usepackage[nolist]{acronym}
\usepackage{mathtools}
\usepackage{multirow}
\usepackage{scalerel}
\usepackage{tikz}
\usetikzlibrary{svg.path}

\definecolor{mcolor}{RGB}{255,10,25}

\defcitealias{Bellardini21}{B21}

\definecolor{orcidlogocol}{HTML}{A6CE39}
\tikzset{
  orcidlogo/.pic={
    \fill[orcidlogocol] svg{M256,128c0,70.7-57.3,128-128,128C57.3,256,0,198.7,0,128C0,57.3,57.3,0,128,0C198.7,0,256,57.3,256,128z};
    \fill[white] svg{M86.3,186.2H70.9V79.1h15.4v48.4V186.2z}
                 svg{M108.9,79.1h41.6c39.6,0,57,28.3,57,53.6c0,27.5-21.5,53.6-56.8,53.6h-41.8V79.1z M124.3,172.4h24.5c34.9,0,42.9-26.5,42.9-39.7c0-21.5-13.7-39.7-43.7-39.7h-23.7V172.4z}
                 svg{M88.7,56.8c0,5.5-4.5,10.1-10.1,10.1c-5.6,0-10.1-4.6-10.1-10.1c0-5.6,4.5-10.1,10.1-10.1C84.2,46.7,88.7,51.3,88.7,56.8z};
  }
}
\newcommand\orcidicon[1]{\href{https://orcid.org/#1}{\mbox{\scalerel*{
\begin{tikzpicture}[yscale=-1,transform shape]
\pic{orcidlogo};
\end{tikzpicture}
}{|}}}}

\title[3D stellar abundances in FIRE MW-mass galaxies]{3D elemental abundances of stars at formation across the histories of Milky Way-mass galaxies in the FIRE simulations}

\author[M. A. Bellardini]{
Matthew A. Bellardini\orcidicon{0000-0002-5663-207X}$^{1}$\thanks{E-mail: mbellardini@ucdavis.edu}, Andrew Wetzel\orcidicon{0000-0003-0603-8942}$^{1}$, Sarah R. Loebman\orcidicon{0000-0003-3217-5967}$^{2}$, and Jeremy Bailin\orcidicon{0000-0001-6380-010X}$^{3}$
\\
$^{1}$Department of Physics \& Astronomy, University of California, Davis, CA 95616, USA\\
$^{2}$Department of Physics, University of California, Merced, 5200 Lake Road, Merced, CA 95343, USA\\
$^{3}$Department of Physics and Astronomy, University of Alabama, Box 870324, Tuscaloosa, AL 35487-0324, USA
}

\date{Accepted XXX. Received YYY; in original form ZZZ}

\pubyear{2022}

\begin{document}
\label{firstpage}
\pagerange{\pageref{firstpage}--\pageref{lastpage}}
\maketitle

\begin{acronym}
\newacro{MW}{Milky Way}
\newacro{GMC}{giant molecular clouds}
\newacro{LG}{Local Group}
\newacro{SN}{supernova}
\newacro{MDF}{metallicity distribution function}
\newacro{ISM}{interstellar medium}
\newacro{FIRE}{Feedback In Realistic Environments}
\newacro{DM}{Dark Matter}
\newacro{MFM}{Meshless Finite Mass}
\newacro{AMR}{adaptive mesh refinement}
\end{acronym}

\begin{abstract}
We characterize the 3-D spatial variations of \FeH, \MgH, and \MgFe{} in stars at the time of their formation, across $11$ simulated \ac{MW}- and M31-mass galaxies in the FIRE-2 simulations, to inform initial conditions for chemical tagging.
The overall scatter in \FeH{} within a galaxy decreased with time until $\approx 7 \Gyr$ ago, after which it increased to today: this arises from a competition between a reduction of azimuthal scatter and a steepening of the radial gradient in abundance over time.
The radial gradient is generally negative, and it steepened over time from an initially flat gradient $\gtrsim 12 \Gyr$ ago.
The strength of the present-day abundance gradient does not correlate with when the disk `settled'; instead, it best correlates with the radial velocity dispersion within the galaxy.
The strength of azimuthal variation is nearly independent of radius, and the $360$ degree scatter decreased over time, from $\lesssim 0.17 \dex$ at $t_{\rm lb} = 11.6 \Gyr$ to $\sim 0.04 \dex$ at present-day.
Consequently, stars at $t_{\rm lb} \gtrsim 8 \Gyr$ formed in a disk with primarily azimuthal scatter in abundances.
All stars formed in a vertically homogeneous disk, $\Delta$\FeH$\leq 0.02 \dex$ within $1 \kpc$ of the galactic midplane, with the exception of the young stars in the inner $\approx 4 \kpc$ at $z \sim 0$.
These results generally agree with our previous analysis of gas-phase elemental abundances, which reinforces the importance of cosmological disk evolution and azimuthal scatter in the context of stellar chemical tagging.
We provide analytic fits to our results for use in chemical-tagging analyses.
\end{abstract}




\section{Introduction}
\label{sec:intro}

Accurate models to describe the formation of the \ac{MW} are crucial for interpreting and guiding observations of it. Current observational surveys, such as GALactic Archaeology with Hermes \citep[GALAH;][]{DeSilva15, Buder18}, Gaia-ESO \citep[][]{Gilmore12}, the Large Area Multi-Object Fiber Spectroscopic Telescope \citep[LAMOST;][]{Cui12}, and the Apache Point Galactic Evolution Experiment \citep[APOGEE;][]{Majewski17, Ahumada20, Jonsson20}, have measured abundances for hundreds of thousands of stars, and future surveys, including the Sloan Digital Sky Survey V \citep[SDSS-V;][]{Kollmeier17}, 4-metre Multi-Object Spectrograph Telescope \citep[4MOST;][]{deJong19}, the WHT Enhanced Area Velocity Explorer \citep[WEAVE;][]{Dalton12}, and the MaunaKea Spectroscopic Explorer \citep[MSE;][]{MSE19} will extend the number of spectroscopically observed stars to the millions. These data, combined with high-fidelity models of galactic elemental enrichment, can offer tremendous insight into the formation history of the MW via `chemical-tagging' \citep[][]{Freeman02}.

Chemical tagging is a technique that leverages the elemental abundances of stars as an invariant to connect present-day observations of stars with their birth locations and times. By contrast, stellar orbital parameters change with time from mergers, accretion, and other dynamical scattering processes \citep[for example][]{SB02, Brook04, Roskar08a, SB09a, Loebman11}.

One can consider chemical tagging in two regimes. `Strong' chemical tagging associates stars with their birth cluster \citep[for example][]{PJ20}, while `weak' chemical tagging associates stars with their general birth time and location within the galaxy \citep[for example][]{Wojno16, Anders17}. Crucially, either form of chemical tagging relies on assumptions about the evolution of the spatial distribution of elemental abundances within a galaxy.

In the case of strong chemical tagging, gas clouds from which stars form must be sufficiently internally homogeneous, and the elemental abundances of individual gas clouds must be sufficiently unique from one another. Observations of star clusters indicate this first assumption is valid \citep[for example][]{Ting12, Bovy16}. The extent to which the second requirement is met is less certain: observations of the \ac{MW} and external galaxies indicate radial and azimuthal variations \citep[for example][]{SanchezMenguiano16, Molla19b, Wenger19, Kreckel20} that could represent sufficiently unique abundances of star clusters.

Weak chemical tagging relies on similar assumptions applied instead to larger regions of the disk, for example, different stars at a given radius have homogeneous abundances that are elementally distinct from stars at other radii. In the extreme limits, one could imagine an elementally homogeneous disk, or an extremely clumpy disk in which all star clusters have unique abundances. The former would provide no spatially discriminating power, whereas the latter would in principle provide complete birth information, but would require complex models.

Critically, chemical tagging techniques rely on accurately modeling the evolution of the spatial scale of elemental abundance homogeneity of stars at birth. This provides constraints on the precision with which one chemically can tag stars. \citet{BKF10} previously explored this with a toy model and showed that most star clusters with masses below $\sim 10^5$ are internally homogeneous, but more work is needed to address the local and global degree of elemental abundance homogeneity in the \ac{MW}.

Many works have measured the present abundance variations of stars in the \ac{MW}. These observations indicate that the \ac{MW} stellar disk has a negative vertical gradient, with more meta-rich stars closer to the disk midplane \citep[for example][]{Cheng12, Carrell12, Boeche14, Hayden14}, although the exact magnitude of the gradient varies between observations and exhibits radial dependence \citep{Hayden14}. Additionally, a multitude of observations show that the stellar disk of the \ac{MW} has a negative radial gradient in abundances \citep[for example][]{Boeche13, Boeche14, Anders14, Mikolaitis14, Donor18, Donor20}. However, the measured radial gradient varies significantly between observations, as well as varying with distance from the galactic midplane \citep[for example][]{Boeche14, Hayden14, Mikolaitis14}. Furthermore, \citet{WL19} showed that the magnitude of radial and vertical gradients is sensitive to stellar age.

Understanding how these variations change across cosmic time is imperative for chemical tagging. An often assumed consequence of `inside-out' galaxy formation \citep{MF89, Bird13} is that, at larger lookback times, stellar disk radial gradients were steeper, and they have flattened with decreasing lookback time. This comes (naively) from assuming the strength of the abundance gradient necessarily follows from the strength of the overall surface-density gradient. Observations of mono-abundance stellar populations in the MW typically find older stellar populations have shallower metallicity gradients \citep[e.g.][]{Anders17, Vickers21}, which these authors attribute to radial redistribution processes flattening the gradients of the oldest populations.  However, another possibility is that older stars formed when the MW had a shallower radial gradient, that is, the MW's gradient has steepened over time.

The abundances of stars at formation trace that of gas, and both theoretical models and observations suggest that gradients steepen with time, though with significant uncertainty \citep[see][and references therein]{Molla19a}. Analyzing the FIRE-1 and FIRE-2 cosmological simulations of \ac{MW}-mass galaxies \citet{Ma17} and Bellardini et al.\ (\citeyear{Bellardini21}, hereafter B21) found that gas-phase abundance gradients steepen with decreasing redshift.
The more analytic model for the evolution of gas-phase metallicity gradients by \citet{Sharda21} also indicates gradients tend to steepen with decreasing redshift, however, they may flatten between redshift $\sim 0.2$ and redshift $0$. Additionally, high-redshift observations indicate gas-phase abundance gradients steepen with decreasing redshift \citep[e.g.][]{Curti20}.

In addition, several observations indicate that young stars show azimuthal variations in abundances across a galaxy \citep[e.g.][]{Luck06, Lemasle08, Pedicelli09}. Looking at B-type stars within $500 \pc$ of the sun, \citet{Nieva12} found a scatter of $\approx 0.05 \dex$ in \OH{}. Recently, using APOGEE data, \citet{Ness21} determined that the median scatter in stellar abundance at fixed radius and time is $0.01 - 0.15 \dex$ for abundances generated via supernovae.  However, in general, the evolution of azimuthal scatter of stellar abundances in the MW and MW-mass galaxies is not well understood, because it requires spatially resolved measurements of stellar abundances at high redshifts of lower-mass galaxies that are analogous to a MW progenitor.

Recent simulation work has emphasized the existence of azimuthal variations at $z = 0$. \citet{Solar20} studied young star particles in $106$ disks from an EAGLE simulation (Ref-L025N0752, initial gas mass resolution of $2.26 \times 10^{5} \Msun$) and found the azimuthal variations lead to a scatter of $\sim 0.12 \dex R^{-1}_{\rm eff}$ in the \OH{} radial gradient. Other simulation work \citep[e.g.][]{Grand16b, DiMatteo16} has shown that azimuthal variations can arise in older stars within spiral galaxies because of streaming motion along non-axisymmetric features like bars and spiral arms.  However, as of yet, cosmological simulations have not characterized the evolution of these azimuthal variations, especially for stars at the time of their formation, to provide context for chemical tagging.

In this paper, we characterize and provide fits for the cosmic evolution of 3-D abundance patterns in newly formed stars as a function of lookback time, to inform the initial conditions of stellar abundance distributions (prior to any post-formation dynamical changes), to inform the precision with which chemically tagging can recover stellar birth location and time.  This builds upon our previous analysis \citepalias{Bellardini21}, where we explored the evolution of the elemental abundance distribution of all gas as a proxy for newly formed stars.

\begin{table}
\centering
\caption{
Stellar properties at $z = 0$ of our FIRE-2 MW/M31-mass galaxies. The first column lists the name of each simulation, the second column lists the stellar mass of the disk \citepalias[see][]{Bellardini21}. The third and fourth columns list the cylindrical $R^{*}_{90}$ (the cylindrical radius which includes $90$\% of the stellar mass) calculated using all stars and young (age $< 250 \Myr$) stars, respectively, within a spherical aperture of radius $< 30 \kpc$ at $z = 0$. The fifth column lists the percentage of ex-situ stars, which we define to be stars that formed beyond a spherical radius $r > 30 \kpc$ comoving. The publication that introduced each simulation is: \citet[][]{Hopkins18}$^1$, \citet[][]{Garrison-Kimmel19b}$^2$, \citet[][]{Garrison-Kimmel19a}$^3$, \citet[][]{Garrison-Kimmel17}$^4$, \citet[][]{Wetzel16}$^5$.
}
\begin{tabular}{l|cccc}
\toprule
\thead{Simulation} &
\thead{$M^{\rm star}_{90}$\vspace{1 mm}\\ $[10^{10} \Msun]$} &
\thead{$R^{\rm star,all}_{90}$\vspace{1 mm}\\ $[\kpc]$} &
\thead{$R^{\rm star,young}_{90}$\vspace{1 mm}\\ $[\kpc]$} &
\thead{Ex-situ\\ percent} \\
\midrule
m12m$^1$ & 10.0 & 11.9 & 12.7 & 6.8 \\
Romulus$^2$ & 8.0 & 14.8 & 16.9 & 8.2 \\
m12b$^3$ & 7.3 & 9.2 & 11.6 & 4.7 \\
m12f$^4$ & 6.9 & 13.5 & 17.0 & 3.0 \\
Thelma$^3$ & 6.3 & 11.7 & 15.1 & 3.2 \\
Romeo$^3$ & 5.9 & 14.2 & 16.8 & 2.9 \\
m12i$^5$ & 5.3 & 10.1 & 12.7 & 2.2 \\
m12c$^3$ & 5.1 & 9.2 & 11.8 & 5.2 \\
Remus$^2$ & 4.0 & 12.4 & 16.2 & 5.7 \\
Juliet$^3$ & 3.3 & 9.5 & 16.0 & 2.8 \\
Louise$^3$ & 2.3 & 12.6 & 17.3 & 1.9 \\
\midrule
Mean & 5.9 & 11.7 & 14.9 & 4.2 \\
\bottomrule
\end{tabular}
\label{table:general_host_properties}
\end{table}

\section{Methods}
\label{sec:methods}

\subsection{FIRE-2 Simulations}
\label{subsec:sims}

\begin{figure}
    \centering
    \includegraphics[width = .9\columnwidth]{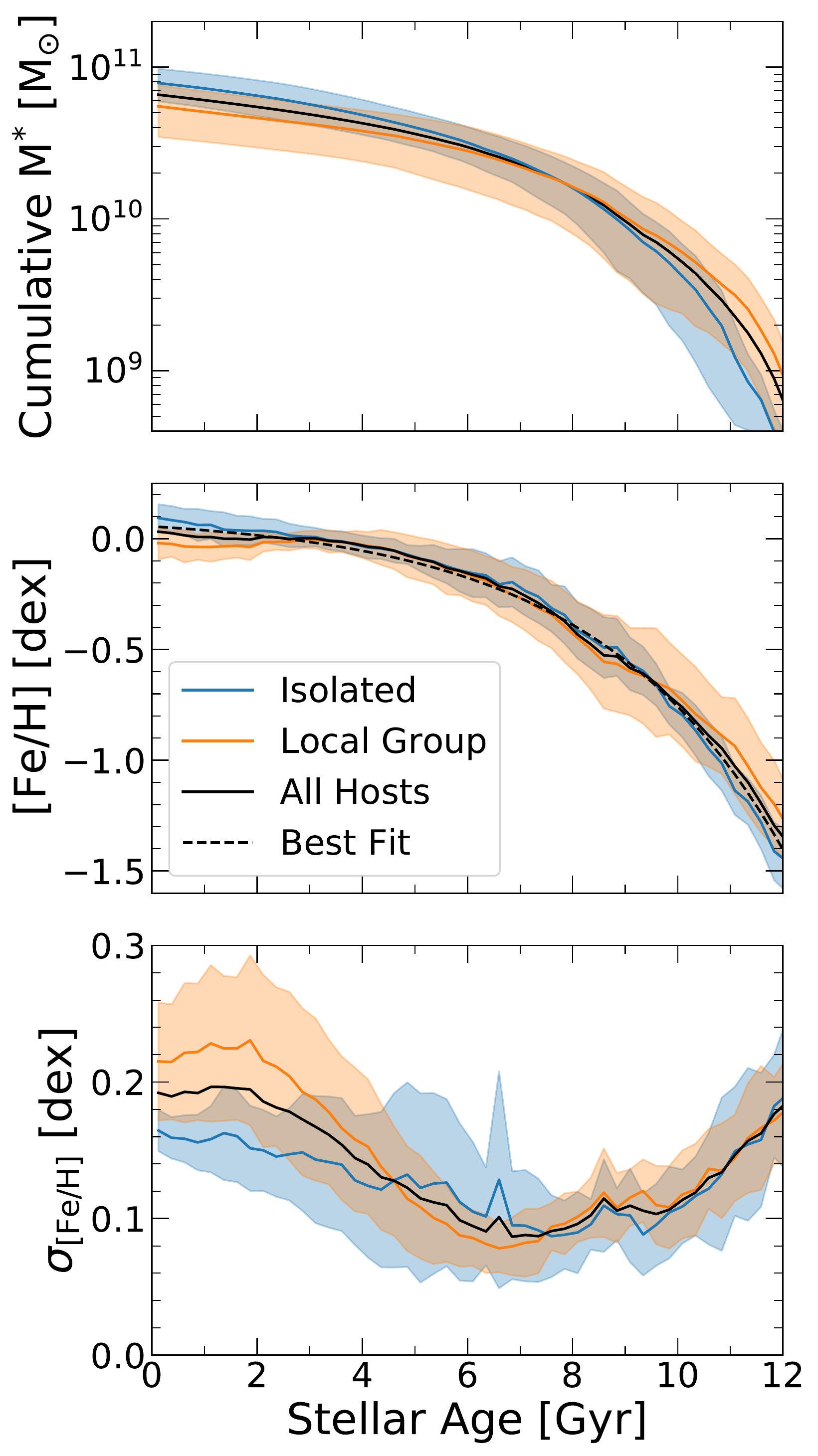}
    \vspace*{0 mm}
    \caption{
    Formation histories of our 11 MW/M31-mass galaxies.
    We include all `in-situ' stars that formed within a spherical radius $r < 30 \times a \kpc$ that remain within a geometrically defined disk (cylindrical radius $R < 20 \kpc$ and vertical height $|Z| < 3 \kpc$) at $z = 0$. We show the average across isolated galaxies (blue), LG-like galaxies (orange), and all galaxies (black).
    We also show the $1-\sigma$ scatter as shaded regions.
    \textbf{Top:} Cumulative stellar mass formed. While the isolated galaxies end up at slightly higher stellar mass at $z = 0$ (on average), the LG-like galaxies formed systematically earlier, as explored in \citet{Santistevan20}.
    \textbf{Middle:} Stellar \FeH, as a proxy for overall metallicity. The black dashed line shows our best fit (see Section~\ref{subsec:functional_forms}). \FeH{} increased until it saturated at $\approx 0.03 \dex$, although the full galaxy sample spans $-0.1 - 0.20 \dex$ at $z = 0$. The LG-like hosts experienced more rapid \FeH{} enrichment at early times, with some reaching $\FeH = -0.5$ $\approx 10 \Gyr$ ago.
    \textbf{Bottom:} The scatter in \FeH{} across the entire galaxy. The scatter at fixed age (time) was high for the oldest stars, then decreased over time down to a minimum $\approx 7 \Gyr$ ago, after which it increased again to $z = 0$. Competition between decreasing bursty/clumpy star formation and increasing steepness of the radial gradient in \FeH{} drive this shape.
    }
    \label{fig:galaxy_property_evolution}
\end{figure}

\begin{figure}
	\includegraphics[width = .91 \columnwidth]{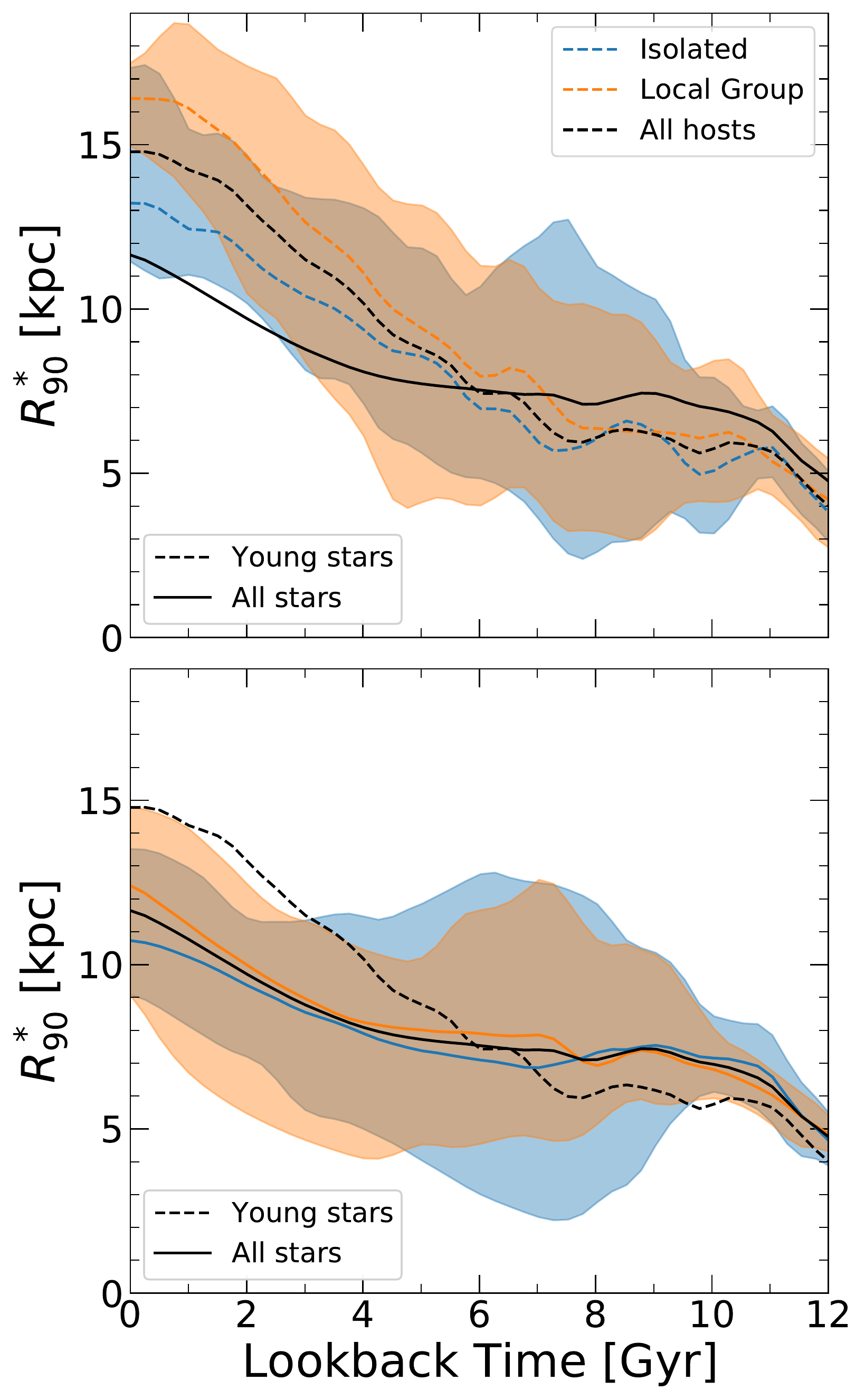}
	\vspace*{-3.5 mm}
    \caption{
    The stellar radius, which encloses 90\% of the stars within $r < 30 \kpc$ comoving ($R^{*}_{90}$),  of the galaxies across time.
    We smooth all lines with a Gaussian filter with $\sigma = 250 \Myr$.
    The dashed lines show $R^{*}_{90}$ using only newly formed stars (ages $< 250 \Myr$), and the solid lines show $R^{*}_{90}$ for all stars.
    The lines show the mean $R^{*}_{90}$ for isolated (blue), LG-like (orange), and all galaxies (black), with the shaded regions showing the range of the full distribution across our sample. $R^{*}_{90}$ grew larger with time, reflecting inside-out radial growth. Prior to $\approx 7.5 \Gyr$ ago, the size of the disks as determined by young versus all stars were similar. After that, $R^{*}_{90}$ for young stars is larger than for all stars.  Additionally, the sizes of disks of young stars in the LG-like galaxies are, on average, larger than those of isolated galaxies at late times.
    }
    \label{fig:size_v_age}
\end{figure}

We use two suites of cosmological zoom-in simulations from the \ac{FIRE} project\footnote{FIRE project web site: \href{http://fire.northwestern.edu}{http://fire.northwestern.edu}} \citep{Hopkins18}. We use $5$ MW/M31-mass galaxies from the \textit{Latte} suite \citep[introduced in][]{Wetzel16}, which have halo masses $ M_{\rm 200m} = 1-2 \times 10^{12} \Msun$, where $M_{\rm 200m}$ refers to the total mass within the radius within which the mean density is $200$ times the mean matter density of the universe. The initial baryon particle mass in the simulations is $7070 \Msun$ (however stellar mass loss leads to star particles having masses of $\approx 5000 \Msun$ at $z = 0$), and the dark-matter mass resolution is $3.5 \times 10^5 \Msun$. Star and dark-matter particles have fixed gravitational force softenings (comoving at $z > 9$ and physical at $z < 9$) with a Plummer equivalent of $\epsilon_{\rm star} = 4 \pc$ and $\epsilon_{\rm dm} = 40 \pc$.  Gas cells have fully adaptive softening, which matches the hydrodynamic kernel smoothing, reaching a minimum softening length of $1 \pc$.

We also include $6$ galaxies from the `ELVIS on FIRE' suite of LG-like MW$+$M31 pairs \citep{ Garrison-Kimmel19a, Garrison-Kimmel19b}. These have mass resolution $\sim 2 \times$ better than \textit{Latte}: the Thelma \& Louise simulation has initial baryon particle masses of $4000 \Msun$ and the Romeo \& Juliet and Romulus \& Remus simulations have initial baryon masses of $3500 \Msun$.

All simulations use the FIRE-2 numerical implementations of star formation, stellar feedback, and fluid dynamics \citep{Hopkins18} using the \ac{MFM} hydrodynamics method of \textsc{Gizmo} \cite[][]{Hopkins15}. \textsc{Gizmo} conserves mass, energy, and momentum of particles to machine accuracy while enabling the adaptive hydrodynamic smoothing of gas elements based on their density.

The FIRE-2 model incorporates physically motivated models of star formation and stellar feedback. 
All simulations include the cosmic ultraviolet background from \citet{Faucher09}.
Gas cells experience metallicity-dependent radiative heating and cooling processes (across a temperature range of $10 - 10^{10}$K) including free-free, photoionization and recombination, Compton, photo-electric and dust collisional, cosmic ray, molecular, metal-line, and fine structure processes, accounting for $11$ elements (H, He, C, N, O, Ne, Mg, Si, S, Ca, Fe). When discussing metallicities in this paper, we scale all elemental abundances to the solar values from \citet{Asplund09}.

Critical for our analysis, the simulations also model the sub-grid diffusion/mixing of elements in gas via turbulent eddies \citep{Su17, Escala18, Hopkins18}. \citetalias{Bellardini21} showed that the details of this implementation strongly affect small scale azimuthal abundance homogeneity.  However, large scale azimuthal abundance variations, as well as vertical and radial trends are largely independent of the strength of our the sub-grid diffusion.

Stars form from gas that is self-gravitating, Jeans-unstable, cold ($T < 10^4$K), and molecular \citep[following][]{KG11}. A newly formed star particle inherits the mass and elemental abundances of its progenitor gas cell. Each star particle represents a single stellar population, assuming a \citet{Kroupa01} stellar initial mass function, which evolves along standard stellar population models. We model time-resolved stellar feedback processes such as continuous mass loss from stellar winds, core-collapse and Ia supernovae, radiation pressure, photoionization, and photo-electric heating. We follow a combination of models \citep{vandenHoek97, Marigo01, Izzard04} synthesized in \citet{Wiersma09} to model stellar winds and their yields. The rates of core-collapse and Ia supernovae come from \textsc{Starburst99} \citep[][]{Leitherer99} and \citet{Mannucci06} respectively. FIRE-2 nucleosynthetic yields follow \citet{Nomoto06} for core-collapse and \citet{Iwamoto99} for Ia supernovae.

We generated cosmological zoom-in initial conditions for all simulations embedded within cosmological boxes with side length $70.4 - 172 \Mpc$, at $z \approx 99$ using \textsc{MUSIC} \citep[][]{Hahn11}. The simulations assume flat $\Lambda$CDM cosmology with parameters broadly consistent with the \citet{Planck20}: $h = 0.68 - 0.71$, $\Omega_{\Lambda} = 0.69 - 0.734$, $\Omega_{\rm m} = 0.266 - 0.31$, $\Omega_{\rm b} = 0.0455 - 0.048$, $\sigma_{8} = 0.801 - 0.82$ and $n_{s} = 0.961 - 0.97$.  For each simulation we save $600$ snapshots from $z = 99$ to $z = 0$, with typical time spacing of $\lesssim 25 \Myr$.

We present the mass, size, and the ex-situ percentage of stars for all galaxies in our analysis in Table~\ref{table:general_host_properties}. We present the mass of all galaxies as measured in \citetalias{Bellardini21}. We show our method for determining the size of the galaxies in Section~\ref{subsec:general_properties}. We define the ex-situ fraction as the fraction of stars currently within the galactic disk (defined geometrically with cylindrical radius $R < 20 \kpc$ and vertical height $|Z| < 3 \kpc$) that formed outside of the spherical aperture of $r = 30 \kpc$ comoving.  We tested using a fixed spherical aperture of $30 \kpc$ rather than a scale factor dependent aperture and found systematically larger disks at $t_{\rm lb} \gtrsim 6 \Gyr$ largely driven by merger activity.  We choose a scale factor dependent initial aperture to remove biases in $R_{90}$ driven by starbursts in merging or close orbiting satellites.

\subsection{Measuring stars at formation}
\label{subsec:properties_at_formation}

We include only stars that end up within a geometrically defined disk at $z = 0$ ($R < 20 \kpc$ and $|Z| < 3 \kpc$).
We present all results in terms of the properties of these stars at the time of their formation, to inform the `initial conditions' for chemical tagging.
Specifically, for each formation time, we look at stars in age bins of width $500 \Myr$ and further subdivide these into $50 \Myr$ age bins.  We then analyze each $50 \Myr$ age bin separately, and average the results of the $10$ bins corresponding to the $500 \Myr$ time window.  This makes no difference to the radial and vertical trends we present, but is important for the azimuthal variations as, given the differential rotational dynamics of the disk, a $500 \Myr$ age bin is larger than an orbital time.

To measure radial abundance profiles we first subdivide the stellar disk into annular bins of width $1 \kpc$ and height $< 1 \kpc$.  For each $50 \Myr$ age bin, within each annular bin, we store the mass-weighted mean stellar abundance.  The total abundance profile is the mass-weighted mean of the profiles in each of 10 age bins making up the full $500 \Myr$ age range.  

To measure vertical abundance profiles, we first define annuli at different radii with width $2 \kpc$.  We subdivide the annuli vertically into slices of height $100 \pc$.  Using the absolute vertical height of star particles, we measure the mass-weighted mean profile in the same way as for the radial abundance profiles.

For the azimuthal abundance variations, we first define annuli at different radii with width $1 \kpc$ and vertical height $< 1 \kpc$.  We analyze the azimuthal scatter at different scales.  For various arclengths we subdivide each annulus into angular bins (the angular bin sizes are defined such that $360^{\circ}$ mod $\phi$ is zero, where $\phi$ is the angular size of the bin in degrees).  The smallest arclength is no smaller than the $1 \kpc$ width of the annulus and the largest arclength is the full annulus. For each $50 \Myr$ age bin we measure the mass-weighted standard deviation of stellar abundances in each angular bin containing at least $3$ star particles.  For each size angular bin we report the average standard deviation as the mass-weighted mean of the standard deviation across all bins, for example, we average 10 bins at $360^{\circ}$, 20 bins at $180 ^{\circ}$, and so on.

We tested varying the minimum number of particles required per annular bin; for both $5$ and $32$ particles there is effectively no difference in the large-scale azimuthal scatter.  However, reducing our minimum number of required particles enables us to measure smaller azimuthal scales. We also experimented with subdividing into different time intervals (25, 50, 100, 250, and 500 Myr intervals) and found that the azimuthal scatter for $50 \Myr$ subdivisions was comparable to the azimuthal scatter for $25 \Myr$ time intervals (the approximate time resolution of snapshots in the simulations) but provided better statistics. Increasing the time interval beyond $50 \Myr$ increased the azimuthal scatter by up to a factor of $\approx 2$ at large lookback times.

\section{Results}
\label{sec:results}

\subsection{Evolution of galaxy stellar mass, metallicity, and size}
\label{subsec:general_properties}

Fig.~\ref{fig:galaxy_property_evolution} shows the average formation histories of our $11$ galaxies versus stellar age.
We include stars within our geometrically defined disk at $z = 0$ that formed `in-situ', that is, within spherical $r < 30\times a \kpc$, where $a$ is the scale factor at the time each star formed, of the main (most massive) progenitor.

Fig.~\ref{fig:galaxy_property_evolution} (top) shows the mean cumulative stellar mass of our $11$ galaxies versus age.  The blue line shows the isolated galaxies and the orange line shows the LG-like galaxies. The solid black line shows the mean across all galaxies. The shaded regions shows the $1-\sigma$ scatter.  The stellar mass increased from a mean of $5.5 \times 10^{8} \Msun$ $12 \Gyr$ ago to a mean of $6.6 \times 10^{10} \Msun$ today. The average mass of the isolated galaxies is slightly ($\approx 1.4 \times$) larger than that of the LG-like galaxies at $z = 0$.
The LG-like galaxies show faster mass growth than the isolated galaxies at early times, in agreement with \citet{Garrison-Kimmel19b} and \citet{Santistevan20}.  At $t_{\rm lb} \gtrsim 8 \Gyr$, the maximum mass difference between the LG-like hosts and the isolated hosts is $\approx 1.9 \times 10^{9} \Msun$.

Fig.~\ref{fig:galaxy_property_evolution} (middle) similarly shows the mean \FeH{} of these galaxies for the same stellar selection.
The dashed black line shows our best fit to the overall mean (see Section~\ref{subsec:functional_forms}). 
The earlier mass assembly of the LG-like galaxies leads to their slightly higher metallicities at earlier times, which is even more pronounced in the upper boundary of the distribution (shaded orange region).
At $10 \Gyr$ ago, the mean $\FeH$ was $\sim -0.75$, with our most enriched (LG-like) galaxy reaching $\FeH = -0.5$ already at that time. This agrees well with observations of old stars in the MW bulge \citep[for example][and references therein]{Bensby17} which find most dwarf stars with \FeH{} $\lesssim -0.5$ are $10 \Gyr$ or older.
However, this metal enrichment largely saturates at late times: stars younger than $\approx 5 \Gyr$ formed with a small range of \FeH{} spanning approximately $-0.09 - 0.03 \dex$.
Overall, this indicates that the \ac{MW}'s LG environment may be key to understanding its early enrichment history.

Fig.~\ref{fig:galaxy_property_evolution} (bottom) shows the average scatter in \FeH{} across the entire galaxy as a function of age. The scatter was larger at early times (that is, for the oldest stars today), and it decreased over time until $\approx 7 \Gyr$ ago. However, after that the scatter increased over time, with the most recently forming stars having again high scatter. As we will explore below, this shape arose from the competition between two processes.
The initial decrease in scatter at early time resulted from the decrease in azimuthal scatter as burstiness of star formation and turbulence in the ISM decreased.
Conversely, the increase in scatter at late times arose from the steepening of the radial gradient.
The epoch of minimum scatter in \FeH{} ($\approx 7 \Gyr$ ago) therefore coincided with the time at which the radial gradient was equal to the azimuthal scatter, that is, when the radial gradient started to overtake the azimuthal scatter as the dominant source of inhomogeneity across the galaxy (see Section~\ref{subsec:azimuthal_v_radial}).

\begin{figure}
    \centering
    \includegraphics[width = 0.91 \columnwidth]{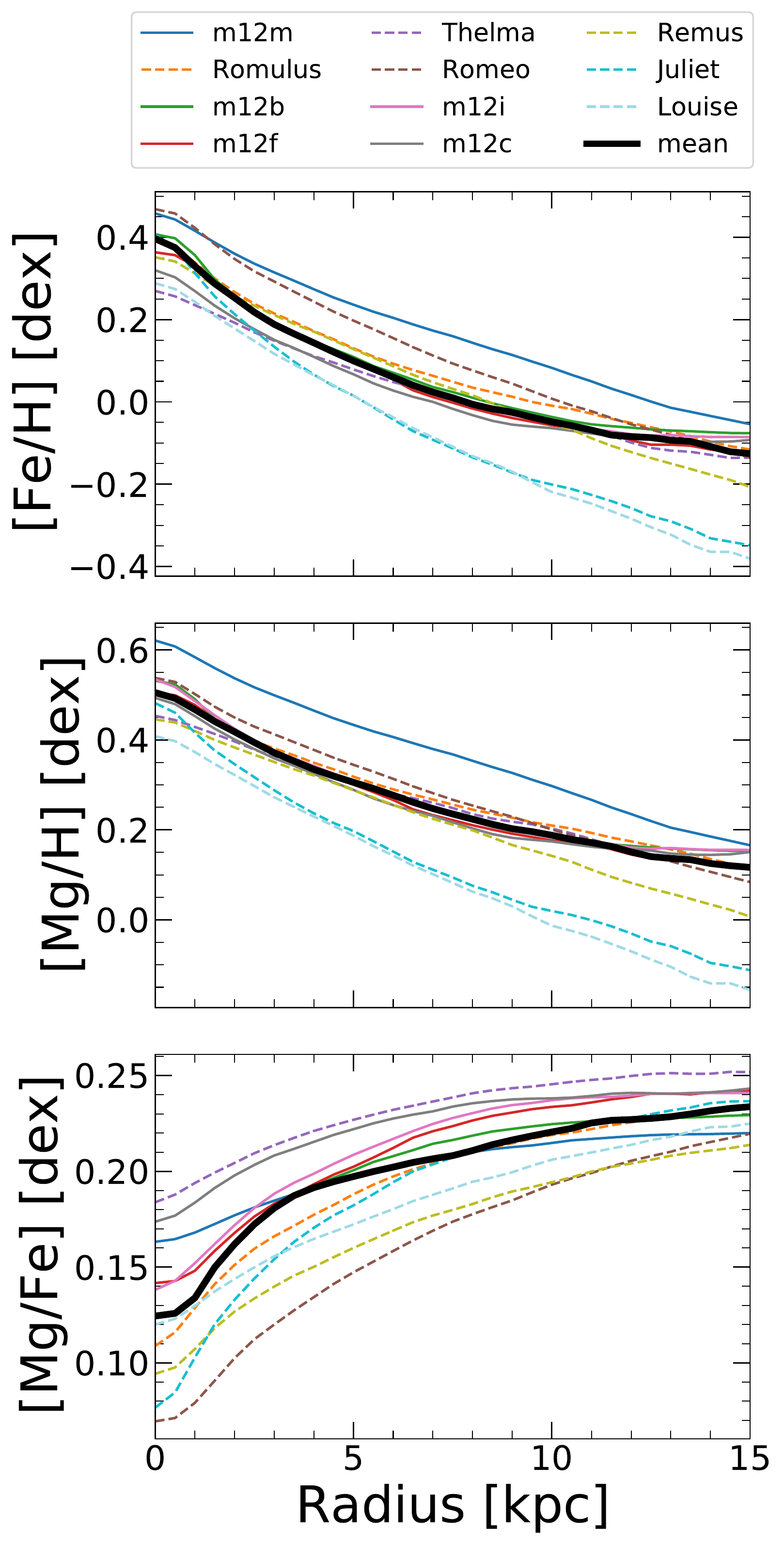}
    \vspace*{-3.5 mm}
    \caption{
    Radial profiles of elemental abundances of stars at formation, for stars that formed within the last $0.5 \Gyr$.
    We include all stars within a vertical height $\pm 1 \kpc$ of the disk plane. 
    We list the galaxies by decreasing stellar mass. The normalization of the \FeH{} and \MgH{} profiles scales roughly with stellar mass. The black line shows the mean, the solid lines show the isolated galaxies, and the dashed lines show the LG-like galaxies; we find no systematic differences between the LG-like and isolated galaxies. \FeH{} and \MgH{} both decrease with radius. The decrease is steeper in the inner region than in the outer region, with the typical transition occurring at $R \approx 5.5 \kpc$. However, \MgFe{} increases with increasing radius, indicating more enrichment from Ia supernovae in the (older) inner disk than in the (younger) outer disk.  This in turn helps to explain the origin of the gradient for young stars as well as the break in the abundance profile at $R \approx 5.5 \kpc$.
    }
    \label{fig:current_stellar_radial_profiles}
\end{figure}

To provide context for our results on radial gradients, Fig.~\ref{fig:size_v_age} shows how the galaxy sizes change with time.
At each time, we fit $R^*_{90}$ simultaneously (iteratively) with $Z^*_{90}$: these two define the radius and height of a cylinder in which the cumulative stellar mass of stars is $90\%$ of total mass of stars within a spherical aperture of $30 \kpc$ comoving.
The top panel shows the size of the disk using only newly formed stars, with ages $< 250 \Myr$, while the bottom panel shows the size of the disk using all stars. Fig.~\ref{fig:size_v_age} shows the mean of the isolated (blue), LG-like (orange), and of all (black) galaxies as well as the full distribution (shaded region) of $R^{*}_{90}$ for the isolated and LG-like hosts.

At lookback times $\gtrsim 5.5 \Gyr$ ago, the disk size as defined by young stars is smaller than the disk size as defined by all stars. However, after this, the disk size as determined by all stars is smaller than that determined by newly formed stars.  This reflects the inside-out radial growth of galaxies: star formation proceeds across larger radii over time \citep[for example][]{Bird13}.

We also note size fluctuations at early times, which are similar to the `breathing mode' fluctuations driven by stellar feedback in low-mass galaxies at $z \sim 0$ \citep{El-Badry16}; however, our smoothing of the sizes partially washes out these short-time trends.
We also measured $R^{*}_{90}$ within a fixed spherical aperture (not scaling with the expansion scale factor), and we found overall similar trends, but with substantial scatter and strong fluctuations at early times, induced by mergers.  Additionally, we tested the robustness of these results to varying the selection region and found no significant difference using a spherical aperture at $z = 0$ rather than a cylindrical `disk-like' region.

For both size metrics, the LG-like galaxies have a larger $R^{*}_{90}$ than the isolated galaxies at $z \sim 0$. The difference is larger for young stars, with the LG-like galaxies having disk sizes $\sim 3 \kpc$ larger than the isolated galaxies ($\sim 1 \kpc$ larger when measuring all stars).
This difference in $R^{*}_{90}$ is consistent with the analysis of a subset of these galaxies (some at lower resolution, initial baryon mass of $2.8 - 3.2 \times 10^{4} \Msun$) in \citet{Garrison-Kimmel18}. We extend this analysis to 6 LG-like galaxies, all at full resolution (initial baryon mass of $3.5 - 4 \times 10^3 \Msun$). This systematic difference implies that some aspect of the LG environment causes stars to form across a larger radius after the onset of disk formation (within the last $\sim 8 \Gyr$), that is, causes more extended disk sizes.
As \citet{Garrison-Kimmel18} discussed, this may relate to stronger gas torques in LG-like environments, and/or this may relate to the earlier formation times of galaxies and their halos in LG-like environments, as \citet{Santistevan20} showed.
We defer a more detailed investigation to future work.

\subsection{Radial abundance profiles at present day}
\label{subsec:present_day_disks}

\begin{figure}
    \centering
    \includegraphics[width=\columnwidth]{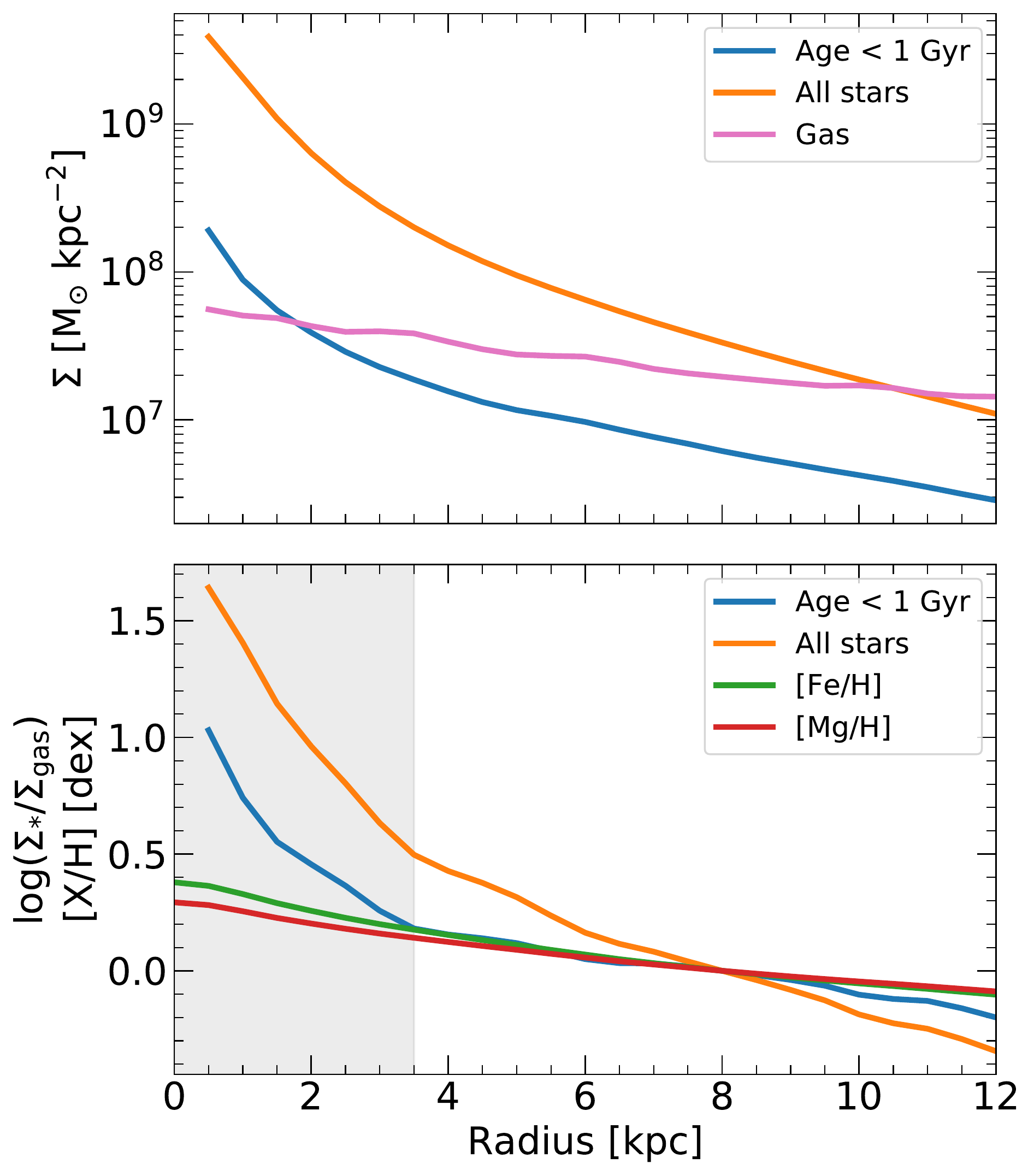}
    \vspace*{-7 mm}
    \caption{\textbf{Top}: Surface density versus radius, for all stars (orange), all gas (pink), and stars younger than $1 \Gyr$ (blue), averaged over our 11 galaxies. The stellar surface densities exhibit steeper inner profiles (in their bulge-like regions) and shallower exponential profiles at larger radii. The gas profile is a shallower exponential at all radii, so gas dominates over stars at $R \gtrsim 10 \kpc$.
    \textbf{Bottom}: The log ratio of stellar to gas surface density, and the \FeH{} and \MgH{} radial abundance, versus radius, all normalized to $R = 8 \kpc$. The ratio of young stars to gas (approximately) should dictate the shape of the abundance profile for newly formed stars.  We find reasonable agreement in the profile shapes at $R \gtrsim 3.5 \kpc$. However, the surface density ratio rises more rapidly than the abundances in the inner few kpc (shaded gray), which may arise from metals being lost to outflows and/or dynamical redistribution of young stars.}
    \label{fig:abundance_and_mass_distributions}
\end{figure}

\begin{figure*}
    \centering
    \includegraphics[width=\linewidth]{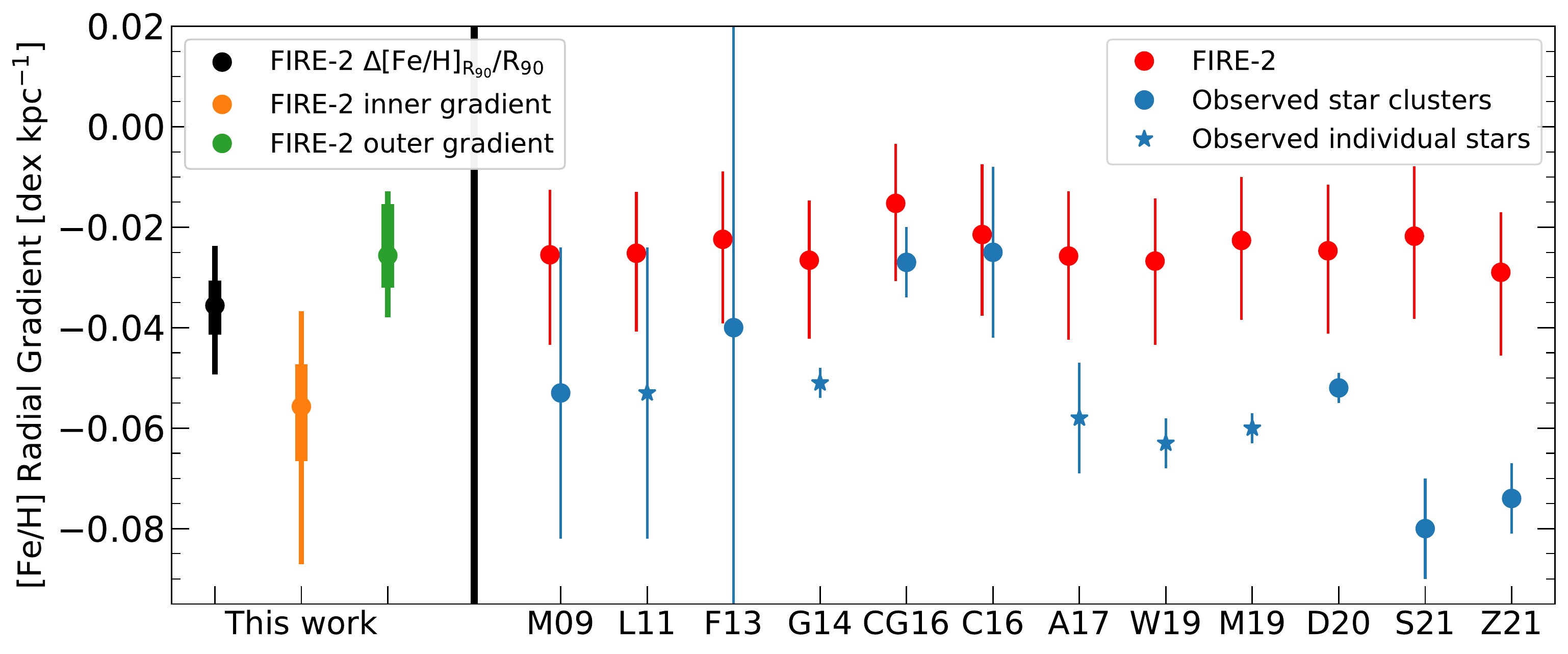}
    \vspace*{-7 mm}
    \caption{
    Radial gradients in \FeH{} for young (ages $< 0.5 \Gyr$) stars at $z = 0$, in the simulations and observed in the Milky Way (MW). For the simulations, our fiducial gradient is a total gradient (black), defined as the difference in metallicity between stars at $R^{*}_{90}$ and at $R = 0 \kpc$ divided by $R^{*}_{90}$. We also show the inner (orange) and outer (green) gradients of the stars measured by fitting a piecewise linear function to the radial profile at $R \leq R^{*}_{90}$.  The thick lines show the 68th percentile of the simulations and the thin lines show the full distribution. We also compare best-fit linear profiles (red) to observational data (blue) of the MW from \citet[][M09]{Magrini09}, \citet[][L11]{Luck11b}, \citet[][F13]{Frinchaboy13}, \citet[][G14]{Genovali14}, \citet[][CG16]{Cantat-Gaudin16}, \citet[][C16]{Cunha16}, \citet[][N16]{Netopil16}, \citet[][A17]{Anders17}, \citet[][W19]{WL19}, \citet[][M19]{Maciel19}, 
    \citet[][D20]{Donor20}, \citet[][S21]{Spina21}, and \citet[][Z21]{Zhang21}.
    Circular and star points show gradients determined via open clusters and individual stars, respectively. For each observational comparison, we fit the linear profiles in the simulations over the same radial range as observed.
    The radial gradients of our simulated galaxies are less steep than most observations of the MW across the same radial range.
    However, the inner gradients in our simulations are more consistent with the MW. As we showed in \citetalias{Bellardini21}, these same simulations are steeper in gas-phase abundance gradients than most nearby MW-mass galaxies.
    }
    \label{fig:sim_vs_obs_grad}
\end{figure*}

We now show results for \FeH, \MgH, and \MgFe. \FeH{} is an easily measured stellar abundance; it is sourced roughly equally by Type Ia supernovae and core-collapse supernovae, however, it is the primary metal produced in Ia supernovae. \MgH{} is our representative $\alpha$ element, because it is the most `pure' $\alpha$ element in the FIRE model, sourced almost entirely via core-collapse supernovae.
\MgFe{} therefore reflects the relative enrichment from core-collapse versus Ia supernovae. We measure Mg as a representative $\alpha$ element rather than O \citepalias[as in][]{Bellardini21}, because it is much easier to measure in stellar atmospheres than O.  Thus, comparison of our results with observations is more straightforward.

Fig.~\ref{fig:current_stellar_radial_profiles} (top two panels) shows the radial profiles of \FeH{} and \MgH{} for stars younger than $500 \Myr$.
Similar to our results for gas-phase abundances \citepalias{Bellardini21}, the galaxy-to-galaxy scatter in normalization primarily reflects the stellar mass-metallicity relationship \citep[for example][]{Tremonti04, Ma16}. These profiles decrease monotonically with radius, because newly formed stars reflect the abundance of the gas, which also decreases with radius, both in these simulations \citepalias{Bellardini21} and in observations \citep[for example][and references therein]{Molla19a}. The decrease in \FeH{} across $0 - 15 \kpc$ is $\approx 0.52 \dex$, nearly identical to the $\sim 0.55 \dex$ decrease in gas \citepalias{Bellardini21}.
The decrease in \MgH{} is weaker, at $\approx 0.39 \dex$. The LG-like galaxies (dashed lines) have marginally steeper gradients in their outer disks. We explore potential causes of this in Section.~\ref{subsec:gradient_correlations} and defer a deeper investigation to future work.

Fig.~\ref{fig:current_stellar_radial_profiles} (bottom) shows that \MgFe{} increases with radius.  Core-collapse supernovae source Mg and Fe relatively equal amounts so decreases in \MgFe{} indicate excess enrichment from type Ia supernovae, which produce more Fe than Mg. The low \MgFe{} in the inner galaxy likely reflects the inside-out radial growth of galaxies \citep[for example][]{Bird21} seen in Fig.~\ref{fig:size_v_age}.
Because stars in the inner galaxy are older on average, the inner galaxy has experienced more enrichment from Ia supernovae, while the outer disk is preferentially more enriched in Mg from core-collapse supernovae.
This also explains why the \FeH{} gradient is steeper than that of \MgH.

Fig.~\ref{fig:current_stellar_radial_profiles} also shows that the gradient is not linear: the slope is typically steeper in the inner region.
We measure a break radius for each galaxy by fitting a two-component piece-wise linear function to each profile. See Fig.~\ref{fig:break_radius_v_mass} for the break radius of each galaxy.

We explore the surface-density profiles in the galaxies to understand better the driver of the break in the abundance profile. Fig.~\ref{fig:abundance_and_mass_distributions} (top) shows the mean surface density profiles for stars younger than $1 \Gyr$ (blue), all stars (orange), and all gas (pink) within $1 \kpc$ of the galactic midplane, at $z = 0$. Fig.~\ref{fig:abundance_and_mass_distributions} (bottom) shows the log of the mean ratio of the stellar to gas surface density for young stars (blue) and all stars (orange), as well as the radial profiles of \FeH{} (green) and \MgH{} (red), all normalized at $R = 8 \kpc$.

The elemental abundance in gas, and therefore in newly formed stars, approximately should scale with the ratio of stellar to gas mass, at least in the limiting case of instantaneous local enrichment. This is not the case in the inner $\sim 3.5 \kpc$ (shaded gray) of the galaxies, which implies metal loss from the inner galaxy. The details of this are beyond the scope of this paper, in which we emphasize the disk component.

The radial change beyond $\sim 3.5 \kpc$ is steeper for all stars than for young stars. This likely explains the shallower radial gradient seen in \MgH{} relative to \FeH. Mg is an $\alpha$ element and is primarily sourced by young stars in core-collapse supernovae, so the gradient in \MgH{} traces the young stellar mass fraction. Fe is sourced roughly equally through core-collapse and Ia supernovae, so \FeH{} traces the stellar mass fraction of older stars on average. Thus a steeper gradient is expected for \FeH{} than for \MgH.

The slope of the ratio of young-star to gas surface density is similar to that of the radial abundance profiles beyond $\sim 3.5 \kpc$.  This similarity indicates that this ratio, to first order, determines the abundance profile.  Thus, this ratio likely partially drives the breaks in the abundance profiles.  However, on an individual galaxy level, the breaks in the abundance profiles and the breaks in the young-star to gas surface density profiles are not always in agreement (see Appendix~\ref{appendix:radial_profile_shapes} for more discussion).

\subsection{Present-day radial profile compared to observations}
\label{subsec:observation_comparison}

Fig.~\ref{fig:sim_vs_obs_grad} (left side) shows the total radial gradient in \FeH{} for young (age $< 0.5 \Gyr$) stars across our 11 galaxies in black, which we define as: the difference in \FeH{} between $R = 0 \kpc$ and $R = R^*_{90}$ divided by $R^*_{90}$:
\begin{equation}
    \frac{\Delta \rm{[Fe/H]_{R^{*}_{90}}}}{R^*_{90}} = \frac{\rm{[Fe/H]}(R = 0) - \rm{[Fe/H]}(R = R^*_{90})}{R^*_{90}}
\end{equation}
Unlike all other results in this paper, here we use the locations of stars at $z = 0$, rather than their formation locations, to compare with observations (notably this changes the gradients by $\lesssim 0.002 \dpk$ on average).
The thick line shows the $1-\sigma$ scatter and the thin line shows the full distribution. The median \FeH{} gradient is $-0.036 \dpk$ with the full range of gradients spanning $-0.049$ to $-0.024 \dpk$.

Fig.~\ref{fig:sim_vs_obs_grad} (left side) also shows the median, $1-\sigma$ scatter, and full distribution of the inner (orange) and outer (green) gradients. We compute these via a linear fit to each profile across $0 - R_{\rm break}$ and $R_{\rm break} - R^{*}_{\rm 90}$, where $R_{\rm break}$ is a free parameter of the fit.
While a two-component piecewise linear function does not fully capture the shape of the abundance profile in all cases, we choose this functional form motivated by observations that find a break in the abundance radial profile \citep[for example][]{Andrievsky04, Sestito08, Magrini09, Pancino10, Frinchaboy13, Hayden14, Korotin14, Maciel19, Zhang21}.  
That said, the majority of observations that find such a break do not separate stars (or star clusters) by age, so the observed breaks might not simply reflect the behavior of stars at formation, but also could be affected by radial redistribution \citep[for example][]{Anders17, Minchev18, Quillen18}.

In our simulations, the gradient is steeper in the inner disk than in the outer disk. This is in contrast to the observational results of \citet{Maciel19}, that the abundance gradients of Cepheids tend to be steeper in the outer disk.  However, our results agree with \citet{Magrini09, Korotin14}, that the abundance gradients of Cepheids are flattest in the outer disk. We also agree with \citet{Eilers22}, who find that the MW's radial gradient in abundance is steeper in the inner disk than the outer disk for RGB stars. The inner \FeH{} gradients span a range of $-0.087$ to $-0.037 \dpk$ with a median of $-0.056 \dpk$ and the outer gradients span a range of $-0.038$ to $-0.013 \dpk$ with a median of $-0.026 \dpk$.

Fig.~\ref{fig:sim_vs_obs_grad} (right side) compares the \FeH{} gradients for young stars in our simulations (red) and observations (blue) of individual stars (star points) and young star clusters (circles). In each case, we measure the gradient within a vertical height $< 1 \kpc$ that spans the same radial range as each observation. The error bars span the full range across our 11 galaxies.

The \FeH{} radial gradients of the young stars in our simulations are less steep than most radial gradients observed in the MW, especially for more recent observations. This agrees with our previous results on gas-phase abundance gradients in those simulations compared with the MW in \citetalias{Bellardini21}.
However, we also found that our gas-phase gradients agree well with M31, and they tend to be steeper than gradients observed in most nearby MW-mass galaxies.
At face value, this suggests that the MW's abundance gradient is unusually steep compared with similar-mass galaxies \citep[for example][]{Boardman20a}.

Observational distance uncertainties could affect these measurements of the MW: \citet{Donor18} found that using different distance catalogs can change observed radial gradients by up to 40\%.
Another possibility that one might consider is that the MW disk settled unusually early compared to nearby galaxies and our simulations.
However, as we show in Section~\ref{subsec:gradient_correlations}, earlier disk settling in our simulations does not correlate significantly with a stronger gradient at $z = 0$.
We do not think that our model for sub-grid diffusion of metals imparts shallowness in our radial gradients, because in \citetalias{Bellardini21} we showed that using a lower diffusion coefficient has little to no impact on our radial gradients. However, if our stellar feedback is too strong, this could drive excess turbulence in the gas, flattening our radial gradients.

In summary, our simulated radial gradients of \FeH{} tend to be shallower than the MW when measured over the same radial range, though our distribution of inner and outer gradients does encompass the full range of observed MW radial gradients.
In other words, we at least recover observed MW gradients within different regions of our galaxies, so any discrepancy may be simply in the location of this radial break.
We also emphasize from \citetalias{Bellardini21} that our gradients are similar to M31 and similar to or steeper than nearby MW-mass galaxies.
In future work we will compare our gradients to the MW across the full range of stellar ages.

\subsection{Evolution of the radial abundance profile}
\label{subsec:radial_evolution}

\begin{figure}
    \centering
    \includegraphics[width = 0.85 \columnwidth]{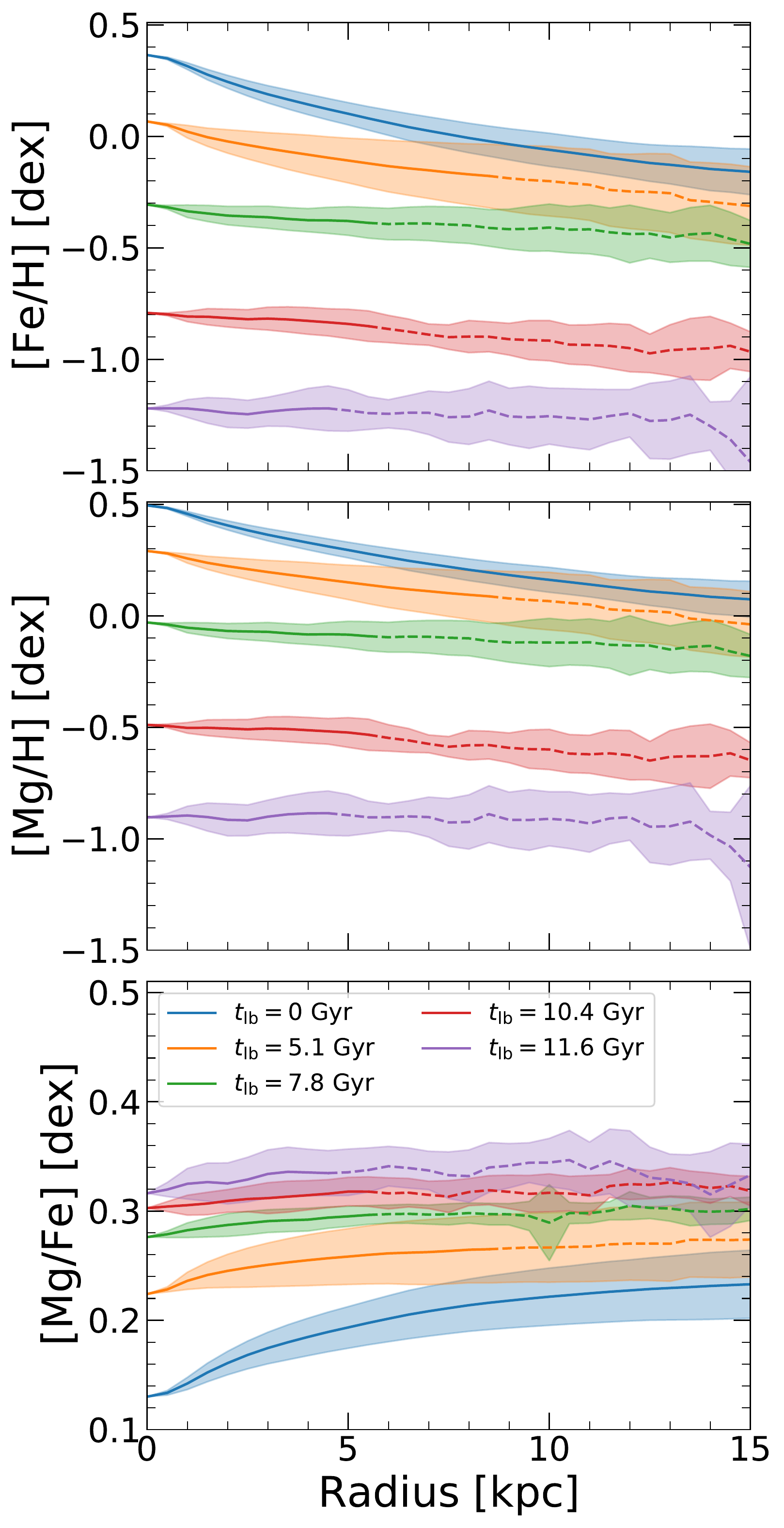}
    \vspace*{-3.5 mm}
    \caption{
    Radial profiles of elemental abundances of stars at formation across our 11 MW/M31-mass galaxies at various lookback times.
    Solid lines shows the profiles out to the average $R^{*}_{90}$ at each lookback time, while the dashed lines show beyond that. The lines show the mean and the shaded regions show the $1-\sigma$ scatter across the galaxies. Because the galaxies had different stellar masses and different abundance normalizations at early times, we re-normalize to the mean abundance at $R = 0$ across the galaxies at each time.
    Similar to the gas-phase abundance profiles of these galaxies \citepalias[see][]{Bellardini21} the stellar abundance profiles are flattest at early times (oldest stars) and steepest at latest times (youngest stars). The combined evolution of the normalization and radial gradients leads to a degeneracy (for chemical tagging), such that stars born at different times and radii can have the same abundance, especially within the last $\sim 8 \Gyr$.
    }
    \label{fig:stellar_radial_profile_evolution}
\end{figure}

\begin{figure}
	\includegraphics[width = .95 \columnwidth]{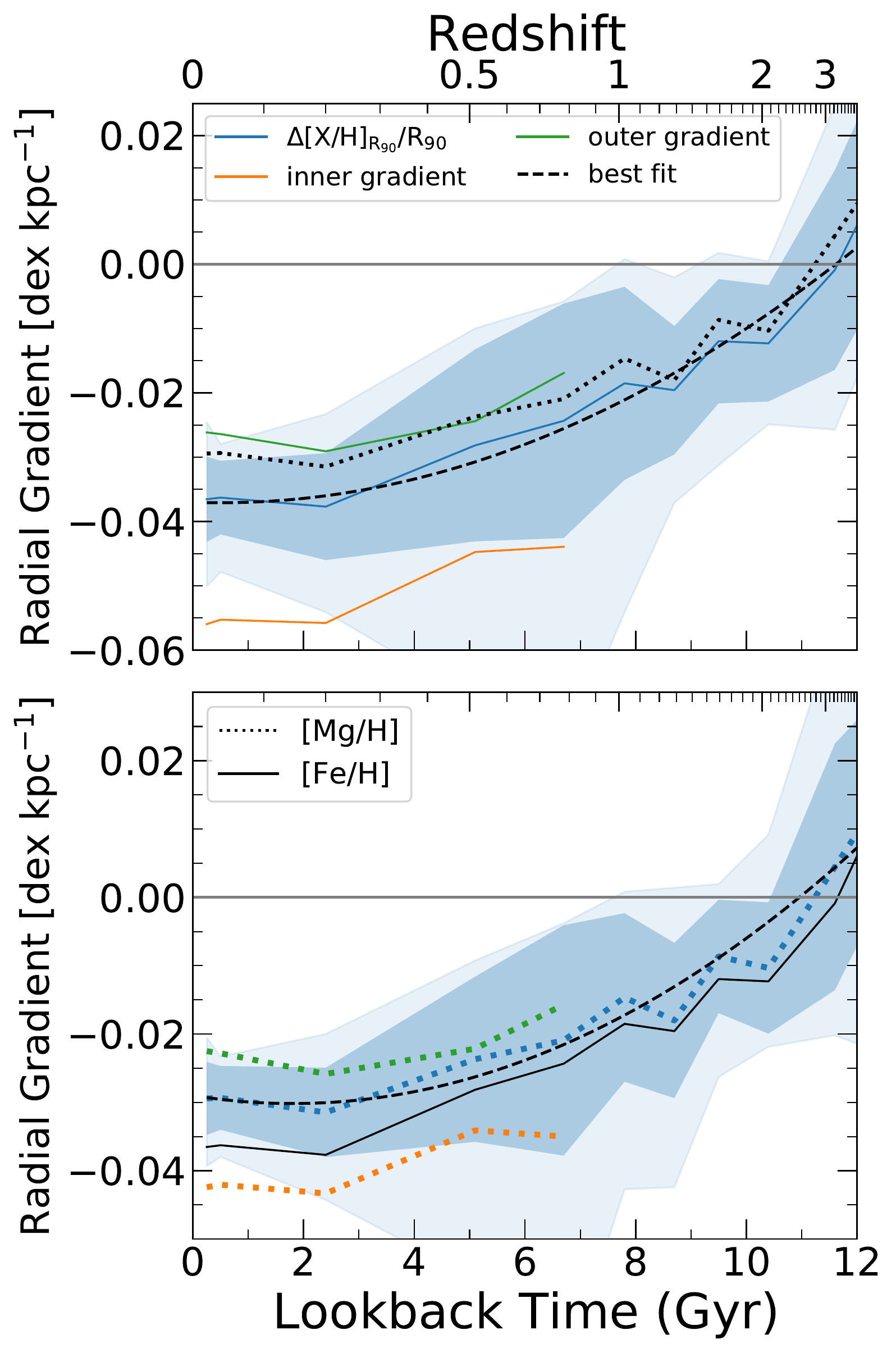}
	\vspace*{-3.5 mm}
    \caption{
    Radial gradients of elemental abundances of stars at formation versus lookback time. The blue line shows the overall gradient, defined as the difference between abundance at $R^{*}_{90}$ and at $R = 0 \kpc$ divided by $R^{*}_{90}$.
    The dark shaded region shows the $1-\sigma$ scatter, and the light shaded regions shows the full distribution across our 11 galaxies.
    The black dashed line shows our best fit to this evolution (see Section~\ref{subsec:functional_forms}).
    The orange and green lines show the inner and outer gradients, respectively, measured by fitting a piece-wise linear function to the abundance profile (see Fig.~\ref{fig:stellar_radial_profile_evolution}), plotted at the lookback times where they reasonably fit the abundance profiles.
    The inner gradient is always steeper than the outer gradient.
    All gradients become steeper over time, and \FeH{} is slightly steeper than \MgH.
    The most extreme gradients in the full distribution reach $\sim -0.077 \dpk$ at $t_{\rm lb} = 6.7 \Gyr$ and $\sim 0.032 \dpk$ at $t_{\rm lb} = 12 \Gyr$ for \FeH{} ($\sim -0.070 \dpk$ at $t_{\rm lb} = 6.7 \Gyr$ and $0.039 \dpk$ at $t_{\rm lb} = 11.6 \Gyr$for \MgH).
    }
    \label{fig:stellar_radial_grad_evolution}
\end{figure}

Fig.~\ref{fig:stellar_radial_profile_evolution} shows the radial profiles of abundances in newly formed stars at different lookback times.  The solid line shows the mean profile at each lookback time out to the average $R^{*}_{90}$, the dashed line shows the profile beyond $R^{*}_{90}$, and the shaded region shows the $1-\sigma$ scatter. Scatter in these galaxies' formation histories, combined with the mass-metallicity relationship, leads to different normalizations of abundances at different times, as in Fig.~\ref{fig:galaxy_property_evolution}, which blurs the trends in the profiles, leading to scatter that primarily reflects different normalizations rather than the shapes of the profiles.
Thus, we normalize the abundance profiles of all galaxies to the average abundance at $R = 0 \kpc$ at each lookback time.

Fig.~\ref{fig:stellar_radial_profile_evolution} (top two panels) shows that, as galaxies evolve, the average metallicity of newly formed stars increased at all radii. The increasing normalization agrees with the trends in Fig.~\ref{fig:galaxy_property_evolution}. Subsequent generations of stars formed from gas that grew more enriched over time.

Beyond this normalization, the shapes of the radial gradients of \FeH{} and \MgH{} for young stars changed with time.  In general the profiles got steeper over time, such that the profiles of stars that formed $\sim 12 \Gyr$ ago were approximately flat, while the most recently formed stars have negative gradients.
These radial profiles of stars at formation trace that of the gas, which \citetalias{Bellardini21} showed grew steeper over time in a similar way, a result of the inside-out radial growth of these galaxies (see Fig.~\ref{fig:size_v_age}).

Fig.~\ref{fig:stellar_radial_profile_evolution} (bottom) shows that \MgFe{} tends to decrease for newly formed stars at all radii with increasing lookback time.
\MgFe{} drops more in the inner galaxy than the outer galaxy, leading to a steepening positive profile over time.  This qualitatively matches the results for gas-phase abundances in \citetalias{Bellardini21}, likely because the older inner galaxy experienced more Ia supernovae, which preferentially enrich the gas (hence the newly forming stars) with Fe, whereas core-collapse supernovae preferentially enrich the younger outer galaxy with $\alpha$ elements like Mg.

Fig.~\ref{fig:stellar_radial_grad_evolution} shows the evolution of radial gradients in \FeH{} and \MgH{} for newly formed stars versus lookback time.  The blue line shows the mean of our fiducial radial gradient: $\Delta \text{[Fe/H]}_{R^{*}_{90}}/R^{*}_{90}$.  The dark blue shaded region shows the $1-\sigma$ scatter and the light blue shaded region shows the full distribution across our 11 galaxies.  The orange and green lines show the inner and outer gradients via fitting a two-component piece-wise linear function (see Section~\ref{subsec:present_day_disks}).  However, we only show the two-component gradients out to $\approx 6.7 \Gyr$, because the functional form is not a good fit when the gradients are sufficiently flat.  The black dashed line in the top panel shows the best fit to the \FeH{} evolution as we describe in Section~\ref{subsec:functional_forms}.

The radial gradients became more negative with decreasing lookback time. The mean total gradient was $0.006 \dpk$ ($0.009 \dpk$) for \FeH{} (\MgH{}) for stars that formed $12 \Gyr$ ago, steepening to $-0.037 \dpk$ ($-0.029 \dpk$) for \FeH{} (\MgH{}) for stars that formed $< 0.5 \Gyr$ ago.  The trend of steeper radial gradients in \FeH{} is consistent at nearly all lookback times. We discuss the functional form of this evolution in Section~\ref{subsec:functional_forms}.

\begin{figure*}
    \centering
    \includegraphics[width=\textwidth]{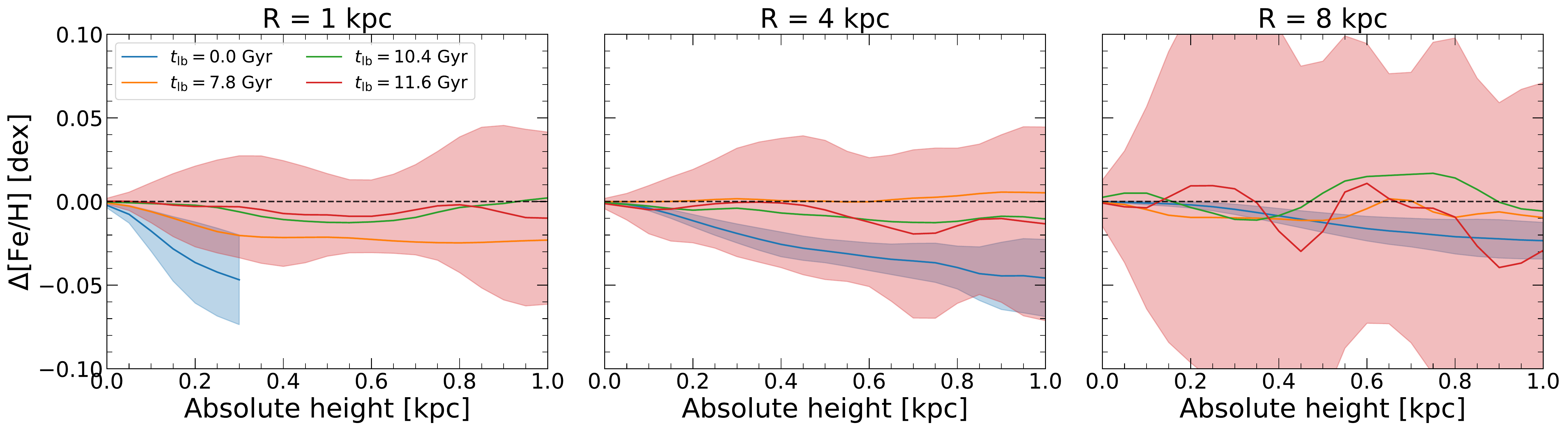}
    \vspace*{-6.5 mm}
    \caption{Vertical profile (relative to the disk mid-plane) of \FeH{} of stars at formation, at $3$ radial annuli centered on $R = 1 \kpc$ (left) $R = 4 \kpc$ (center) and $R = 8 \kpc$ (right) with width $\pm 1 \kpc$. The solid lines show the mean deviation in abundance at each height, and the shaded regions show the $1-\sigma$ scatter across our 11 galaxies. We smooth all lines with a Gaussian filter with $\sigma=0.05\kpc$. \FeH{} is generally invariant with increasing height from the midplane. The only exception is stars at present-day, which show a modest decrease in \FeH{} and with increasing height ($\gtrsim 0.05 \dpk$ slope for all $R \lesssim 4 \kpc$), tracing the abundance pattern of the star-forming gas in this region.}
    \label{fig:stellar_vertical_profile_evolution}
\end{figure*}

Fig.~\ref{fig:stellar_radial_grad_evolution} also shows our two-component fit to the gradients, which we do not show at large lookback times, when the profiles were flat (see Fig.~\ref{fig:stellar_radial_profile_evolution}). However, characterizing the transition of the disk from a single-component to a multi-component profile is beyond the scope of this paper. Similar to our measurement of the total gradient, both the inner and outer gradients steepen over time. For the stars forming at $z \sim 0$, the mean inner gradient is $-0.056 \dpk$ ($-0.042 \dpk$) and the mean outer gradient is $-0.026 \dpk$ ($-0.023 \dpk$), for \FeH{} (\MgH{}).
The inner gradient is consistently steeper than the outer gradient at all times, which, as we argue from Fig.~\ref{fig:abundance_and_mass_distributions}, indicates that the ratio of stellar to gas mass always had a steeper profile in the inner galaxy.

\subsection{Vertical profile}
\label{subsec:vertical_profiles}

Fig.~\ref{fig:stellar_vertical_profile_evolution} shows the mean change in abundance as versus distance from the galactic midplane for newly formed stars within $2 \kpc$ wide bins centered at $3$ radii, $R = 1 \kpc$ (left), $R = 4 \kpc$ (center), and $8 \kpc$ (right), at various lookback times.
We measure this change with respect to the midplane abundance value of each host.  We smooth all profiles with a Gaussian filter (using scipy.ndimage.gaussian\_filter1d with $\sigma = 0.05\kpc$). The solid lines show the mean, and the shaded regions show the $1-\sigma$ scatter across our 11 galaxies.  For clarity we show only the scatter at present day and at the largest lookback time.

The \FeH{} and \MgH{} (not shown) profiles show little to no significant variation with increasing distance from the midplane. The mean profiles show slight deviations up to $1 \kpc$, but any systematic trends are typically smaller than the galaxy-to-galaxy $1-\sigma$ scatter, and/or the strength is typically smaller than the measurement precision of most stellar surveys.
These flat vertical profiles are unsurprising, given the similar gas-phase results we explored in \citetalias{Bellardini21}. We thus conclude that vertical gradients in abundances are generally negligible.

The one exception is young stars at present-day in the inner disk, which have a systematic modest negative vertical gradient ($\gtrsim 0.05 \dpk$ out to approximately $4 \kpc$). This declining metallicity with increasing distance from the midplane agrees with some observations of the MW \citep[for example][]{Katz11, Hayden14, Xiang15, WL19}.
We find that this trend is caused by the star-forming gas, that is, it is not affected by any post-formation stellar dynamics on timescales $\lesssim 20 \Myr$. In particular, unlike at larger radii, the metallicity distribution of star-forming gas at small radii near the midplane is highly non-Gaussian, indicating that enrichment timescale is shorter than the mixing timescale.
In other words, given both the higher star-formation (metal enrichment) density and the stronger gravitational potential in the inner galaxy, gas is not able to mix as efficiently in the vertical direction, leading to enhanced enrichment in the midplane.

We additionally explored the vertical profile of \MgFe{} (not shown here). \MgFe{} increases with distance from the midplane in the inner disk at present-day. Here, both \MgH{} and \FeH{} decrease (slightly) with height, so the stars at larger heights are somewhat less enriched. The \MgH{} gradient is slightly weaker, though, which may result from vertical enrichment being more affected by star-forming winds, enriched preferentially in Mg from core-collapse supernovae.

We caution that these simulations do not include black-hole feedback (see Wellons et al. in prep.), which could affect these vertical gradients in the inner galaxy.

\subsection{Azimuthal scatter}
\label{subsec:azimuthal_scatter}

\begin{figure}
    \includegraphics[width=0.87\columnwidth]{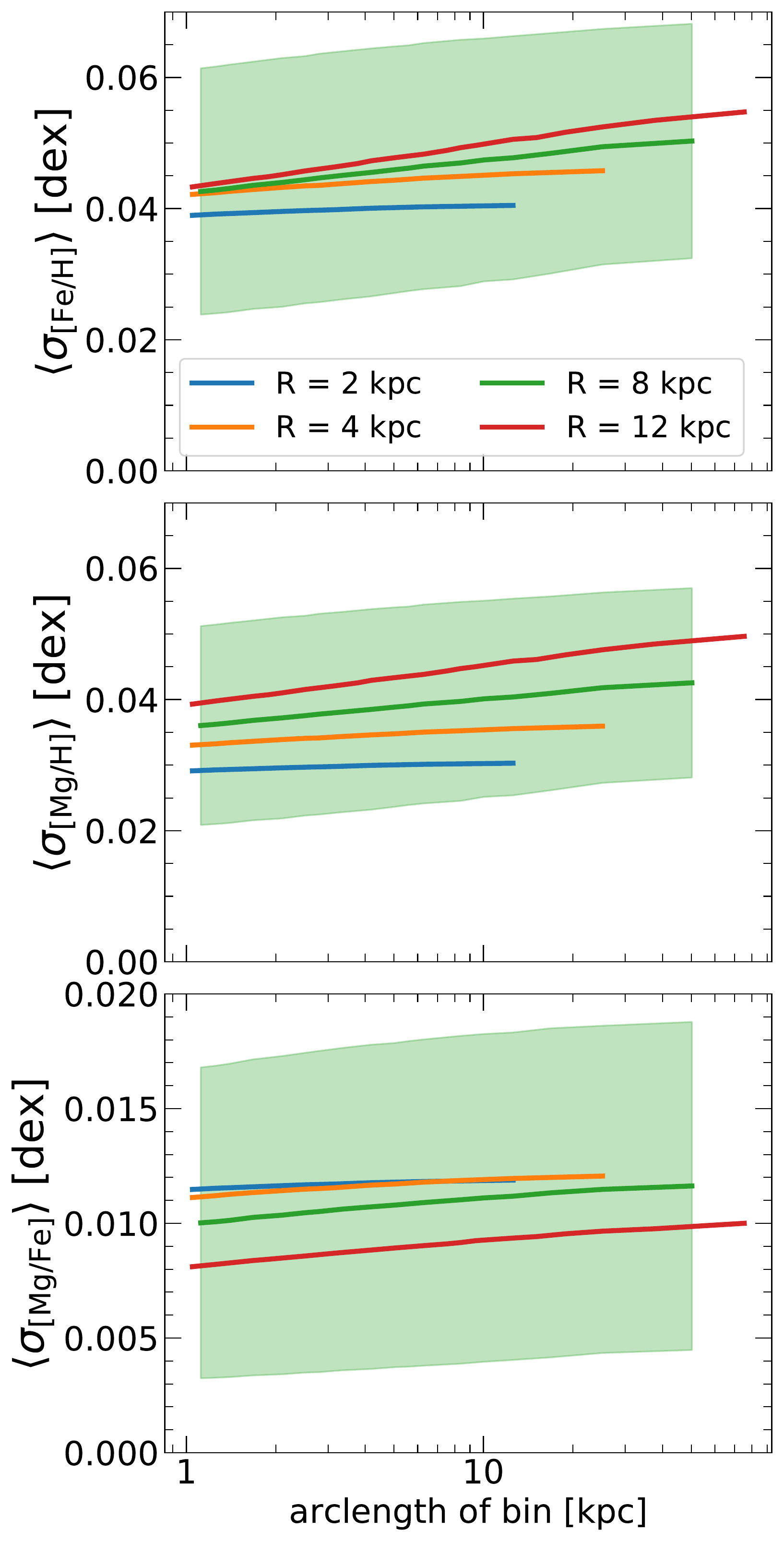}
    \vspace*{-3.5 mm}f
    \caption{
    Azimuthal scatter of elemental abundances versus arclength, for stars at formation within the last $500 \Myr$ at $z = 0$, using annuli $1 \kpc$ in width centered on $4$ different radii, averaged across our 11 galaxies. The shaded region shows the $1-\sigma$ scatter in a fiducial solar cylinder ($R = 8 \kpc$). The azimuthal scatter in \FeH{} and \MgH{} increases with arclength and with radius. The scatter in \MgFe{} decreases slightly with radius. The $360^{\circ}$ scatter in \FeH{} at $R = 8 \kpc$ is $\approx 0.05 \dex$, which agrees well with the scatter in gas in \citetalias{Bellardini21}. The minimal dependence on azimuthal bin width (arclength) agrees well with the dependence at $z = 0$ in \citetalias{Bellardini21}. We do not center these bins on star-forming regions, so even on scales $\sim 1 \kpc$ the scatter remains $\sim 0.03 - 0.04 \dex$.
    }
    \label{fig:young_stellar_azimuthal_scatter}
\end{figure}

\begin{figure}
    \includegraphics[width=0.9\columnwidth]{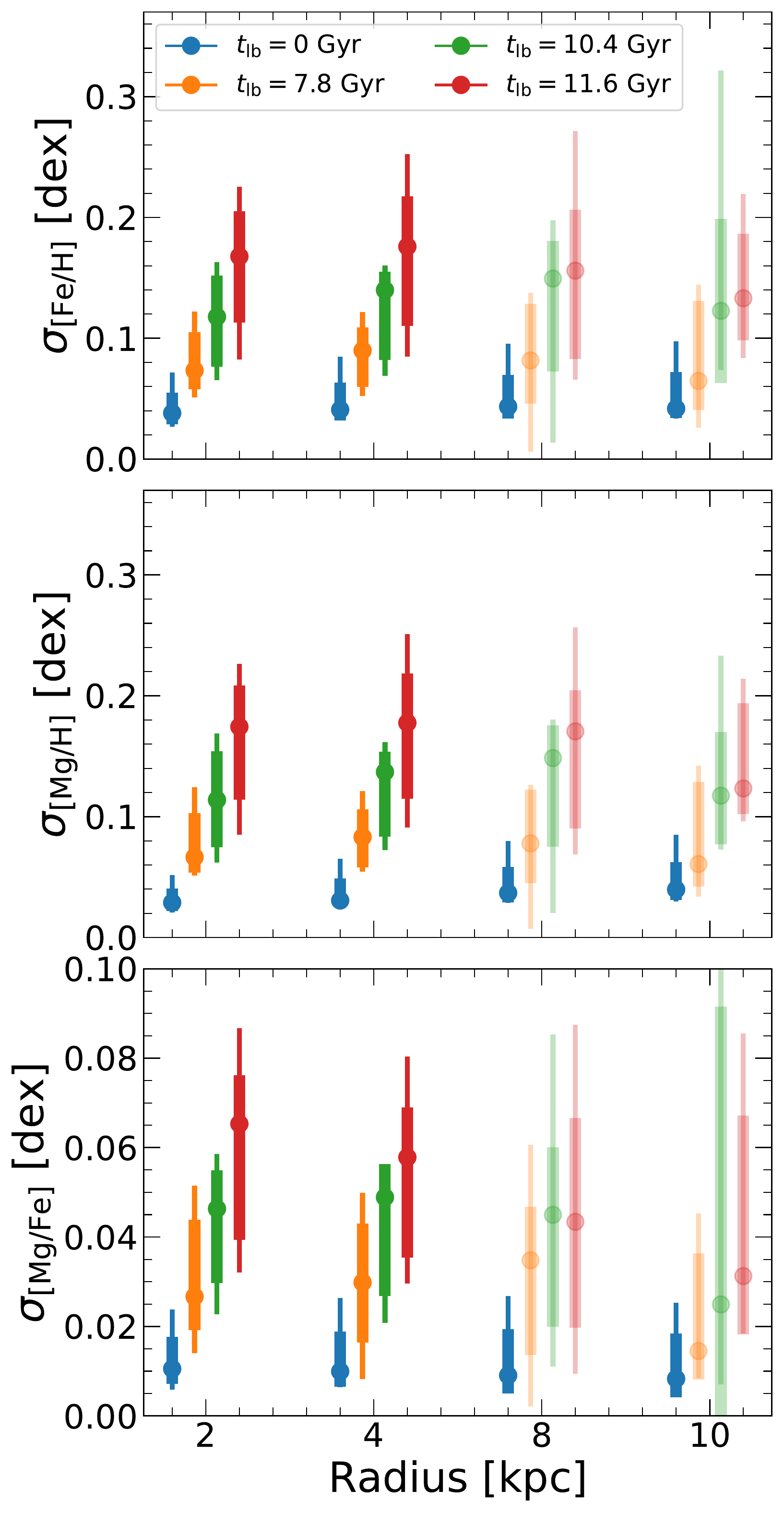}
    \vspace*{-3.5 mm}
    \caption{
    Full ($360^{\circ}$) azimuthal scatter of newly formed stars as a function of radius at various lookback times. The points show the median scatter, the thick lines show the $1-\sigma$ scatter, and the thin lines show the full distribution across our 11 galaxies. The lightly shaded points show the scatter beyond the average $R^{*}_{90}$ of the galaxies. The azimuthal scatter shows little to no dependence on radius. We show only the largest-scale ($360^{\circ}$) azimuthal scatter, because as Fig.~\ref{fig:young_stellar_azimuthal_scatter} shows, the scatter depends minimally on azimuthal bin width. By contrast, the azimuthal scatter increases with lookback time. The median scatter at $8 \kpc$ increases from $\sim 0.04 \dex$ to $\approx 0.16 \dex$ from $t_{\rm lb} = 0 \Gyr$ to $11.6 \Gyr$.  However, this time dependence is weaker than for all gas in these galaxies (see \citetalias{Bellardini21} Fig.7).
    }
    \label{fig:stellar_azimuthal_scatter_evolution}
\end{figure}

\begin{figure*}
    \centering
    \includegraphics[width=0.95\textwidth]{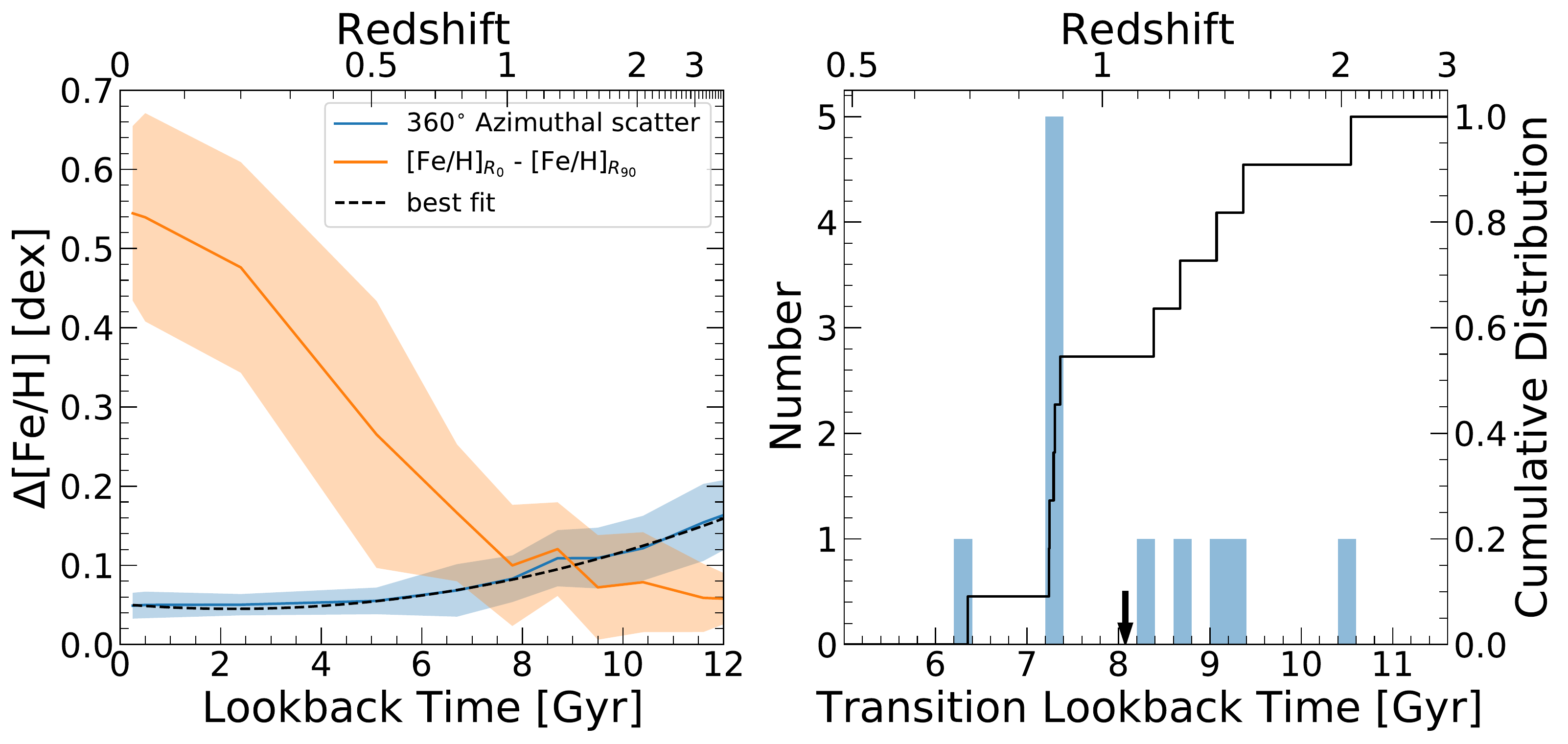}
    \vspace*{-3.5 mm}
    \caption{
    \textbf{Left:} The magnitude of radial versus azimuthal variations in \FeH{} for newly formed stars, versus lookback time.
    The orange line shows the absolute change from $R = 0 \kpc$ to $R^{*}_{90}$, while the blue line shows the $360^{\circ}$ azimuthal scatter. The solid lines show the average and the shaded regions show $1-\sigma$ scatter across our 11 galaxies.
    The intersection of these 2 lines defines a transition age: stars older than this formed in a galaxy dominated by azimuthal scatter, while stars younger than this formed in a disk in which radial variations dominated.
    \textbf{Right:} A histogram of this transition lookback time across our 11 galaxies. The black line shows the cumulative distribution. The black arrow shows the average ($\approx 8 \Gyr$ ago), which is $0.5 - 1 \Gyr$ earlier than the average transition time for gas \citepalias{Bellardini21}.
    }
    \label{fig:azimuthal_scatter_vs_gradient}
\end{figure*}

Fig.~\ref{fig:young_stellar_azimuthal_scatter} shows the azimuthal scatter of elemental abundances at formation of stars at $z = 0$ with ages $< 0.5 \Gyr$ at $4$ radii ($R = 2, 4, 8, 12 \kpc$) as a function of azimuthal arclength.  The solid lines show the mean scatter, and the shaded region highlights the standard deviation for stars that formed at our fiducial solar radius of $R = 8 \kpc$.

The azimuthal scatter at $z = 0$ increases modestly with radius and depends only weakly on azimuthal bin size.  At $R = 2 \kpc$ the mean scatter is essentially independent of the angular bin arclength and the scatter within the whole annulus is $\sim 0.01 \dex$ smaller than the mean scatter in the annulus at $R = 8 \kpc$. At $R = 8 \kpc$ the scatter depends slightly on azimuthal bin width, increasing from $0.043 \dex$ to $0.05 \dex$ for \FeH{} ($0.036 \dex$ to $0.043 \dex$ for \MgH).  Fig.~\ref{fig:young_stellar_azimuthal_scatter} shows that the azimuthal scatter of \MgFe{} decreases slightly with increasing radius, in contrast with the trends for \FeH{} and \MgH{}.  However, the radial dependence is smaller than the host-to-host scatter.  Importantly, we do not center our bins on individual star-forming regions (star clusters), so, for all panels in Fig.~\ref{fig:young_stellar_azimuthal_scatter}, the scatter does not go to $0 \dex$ at small arclengths, as one might expect.

Fig.~\ref{fig:stellar_azimuthal_scatter_evolution} shows the evolution of the azimuthal scatter, specifically, showing the $360^{\circ}$ scatter versus radius at various lookback times.  We show only the $360^{\circ}$ scatter here, because the difference between the scatter at the smallest and largest scales is minimal ($< 0.018 \dex$) for all lookback times and radii.  The points show the median scatter, the thick lines show the $1-\sigma$ scatter, and the thin lines show the full distribution across our 11 galaxies.  The lightly shaded points indicate radii that are larger than the average $R^*_{90}$ at a given lookback time.

The azimuthal scatter at all radii and all azimuthal bin sizes generally increase with increasing lookback time. Stars that have formed more recently formed in a more homogeneous azimuthal disk than stars that formed earlier.  This agrees with the trends for gas abundances in \citetalias{Bellardini21}, though at large lookback times the scatter in newly formed stars is generally smaller than the scatter seen for all gas (see Appendix~\ref{appendix:stars_vs_gas}).
The smaller scatter in stars results in part from stars forming from gas that is preferentially metal rich, especially at large lookback times \citepalias[for example][]{Bellardini21} and at large radii. Additionally, newly formed stars are more spatially clustered than all gas cells.
This explains the weaker (shallower) dependence on azimuthal bin size for young stars than for gas: the fraction of bins containing no mass is larger for stars than for gas, so increasing the binsize does not necessarily include more/different stars per bin.

The azimuthal scatter for stars shows little dependence on radius, in contrast to the azimuthal scatter for gas \citepalias{Bellardini21}. Here, the change in scatter with radius is in general less than $\sim 0.02 \dex$. Thus, stars throughout the galaxy form in essentially equally azimuthally homogeneous conditions.

Fig.~\ref{fig:stellar_azimuthal_scatter_evolution} (bottom) shows the evolution of azimuthal scatter for \MgFe{}.  Similar to gas \citepalias{Bellardini21}, the scatter in \MgFe{} is much smaller at all radii than the scatter in \FeH{} or \MgH.  However, the scatter in \MgFe{} follows the same trend of increasing with increasing lookback time.  The scatter in \MgFe{} shows less dependence on azimuthal bin width than the scatter for \FeH{} and \MgH{} with a maximum change of $\approx 0.006 \dex$ between the smallest and largest scales.  The galaxy-to-galaxy standard deviation is also smaller for \MgFe{} than for \FeH{} and \MgH.

\subsection{Strength of azimuthal versus radial variation}
\label{subsec:azimuthal_v_radial}
Given the first-order approach of chemical evolution/chemical tagging models to neglect azimuthal scatter \citep[for example][]{Minchev18, Molla19a, Frankel20}, it is critical to identify when the assumption of minimal azimuthal scatter is valid.  These models are only accurate representations of the MW when azimuthal abundance variations are smaller than radial abundance variations.  The general increase in the steepness of radial gradients with time (seen in Section.~\ref{subsec:radial_evolution}) and the general decrease in azimuthal scatter with time (seen in Section.~\ref{subsec:azimuthal_scatter}) implies there must be some transition time prior to which azimuthal scatter is the dominate source of abundance variations and after which radial variations dominate.

Fig.~\ref{fig:azimuthal_scatter_vs_gradient} (left) compares the strength of azimuthal scatter to radial abundance change for newly formed stars as a function of lookback time. The orange line shows the mean abundance change in radial abundance between $R = 0 \kpc$ and $R^{*}_{90}$.
The blue line shows the mean $360^{\circ}$ azimuthal scatter of \FeH{}, which we averaged across $R = 2, 4, 6, 8,$ and $10 \kpc$, given the modest radial dependence.
The dashed line shows the best fit to the evolution of the azimuthal scatter (see Section~\ref{subsec:functional_forms}; Fig.~\ref{fig:stellar_radial_profile_evolution} shows the fit to the radial change). The shaded regions show the $1-\sigma$ scatters across our 11 galaxies.

As Fig.~\ref{fig:stellar_radial_grad_evolution} showed, the strength of the radial variations increased with time as the radial gradient steepened. Also, as Fig.~\ref{fig:stellar_azimuthal_scatter_evolution} showed, the azimuthal scatter decreased with time at all radii. The point at which these cross identifies a transition epoch, at which newly formed stars transitioned from forming in a galaxy primarily dominated by azimuthal scatter to a disk primarily dominated by a radial gradient. This transition necessarily correlates with the gas disks transitioning to being rotationally dominated; \citet{Ma17} showed that strong radial gradients are only found in galaxies with a gas disk characterized by well-ordered rotation.

This is a critical transition period to characterize for chemical tagging, because it in effect identifies the maximum age of stars for which 1-D radial models for chemical tagging provide a good approximation. For all stars that formed prior to this transition age, their abundance was influenced more by their azimuthal location than their radius at birth.

Fig.~\ref{fig:azimuthal_scatter_vs_gradient} (right) shows a histogram of this transition time for each simulated galaxy, which spans $\approx 6.4$ to $\approx 10.6 \Gyr$ ago.
The black line shows the cumulative distribution.
The black arrow shows the mean transition time, $\approx 8 \Gyr$ ago.
This transition age for stars is slightly earlier than the transition age for gas \citepalias[as presented in][]{Bellardini21}, which varied with radius from $7.4 \Gyr$ ago at $R = 4 \kpc$ to $6.9 \Gyr$ ago at $R = 12 \kpc$ (see Appendix~\ref{appendix:stars_vs_gas} for more discussion).

\begin{table*}
    \centering
    \caption{The Spearman rank correlation coefficient and corresponding $p$-value between the total stellar \FeH{} gradient or the outer stellar \FeH{} gradient in young stars (age $< 500 \Myr$) and different metrics of the galaxies' formation histories. The metrics we use, ranked by the average p-value of the total and outer gradient correlations, are: the ratio of the radial velocity dispersion of young stars to their average circular velocity ($\sigma^{\rm star, young}_{v_{r}}/v_{\rm circ}$), the radial velocity dispersion of young stars ($\sigma^{\rm star, young}$), the median age of stars in the galaxy, the circularity parameter of young stars $j_{z}/j_{c}$ \citep{Abadi03}, the transition time from bursty to smooth star formation in \citet{Yu21}, the transition time from azimuthal scatter domination to radial gradient domination in Section~\ref{subsec:azimuthal_v_radial}, the break radius of a two-component linear profile fit to the ratio of young stellar surface density to gas surface density shown in Fig.~\ref{fig:break_radius_v_mass}, the change in the surface density ratio of young stars to gas over the same radial range divided by $R_{90} - 3 \kpc$ ($\nabla \Sigma^{\rm star, young} / \Sigma^{\rm gas}$), the $R^{\rm star, all}_{90}$ and $R^{\rm star, young}_{90}$ in Section~\ref{subsec:general_properties}, and the change in the surface density of young stars from $R = 3 \kpc$ to $R_{90}$ divided by $R_{90} - 3 \kpc$ ($\nabla \Sigma^{\rm star, young}$).  We find the only correlations with significant $p$-values to be with: median age, $\sigma^{\rm star, young}$, and $\sigma^{\rm star, young}_{v_{r}}/v_{\rm circ}$.}
    \begin{tabular}{l|rr|rr}
    \toprule
    \multicolumn{1}{c|}{\multirow{2}{*}{Correlation Metric}} & 
    \multicolumn{2}{c|}{Total Gradient} &
    \multicolumn{2}{c}{Outer Gradient} \\
    \multicolumn{1}{c|}{} &
    \multicolumn{1}{c}{Correlation} &
    \multicolumn{1}{c|}{p-value} &
    \multicolumn{1}{c}{Correlation} &
    \multicolumn{1}{c}{p-value} \\
    \midrule
    $\sigma^{\rm star, young}_{v_{r}} / v_{\rm circ}$ & 0.727 & 0.011 & 0.836 & 0.001 \\
    $\sigma^{\rm star, young}_{v_{r}}$ & 0.655 & 0.011 & 0.764 & 0.006 \\
    Median stellar age & -0.745 & 0.008 & -0.682 & 0.021 \\
    \midrule
    $j_{z}/ j_{c}$ & -0.409 & 0.212 & -0.509 & 0.110 \\
    Bursty to smooth SFR time & -0.464 & 0.151 & -0.427 & 0.19 \\
    Transition lookback time & -0.351 & 0.290 & -0.469 & 0.145 \\
    $\Sigma^{\rm star, young} / \Sigma^{\rm gas}$ & 0.127 & 0.709 & 0.336 & 0.312 \\
    $\nabla \left( \Sigma^{\rm star, young} / \Sigma^{\rm gas} \right)$ & 0.027 & 0.937 & -0.291 & 0.385 \\
    $R^{\rm star, all}_{90}$ & 0.223 & 0.509 & -0.036 & 0.916 \\
    $R^{\rm star, young}_{90}$ & -0.005 & 0.989 & -0.196 & 0.564 \\
    $\nabla \Sigma^{\rm star, young}$ & 0.027 & 0.937 & -0.127 & 0.709 \\
\bottomrule
    \end{tabular}
    \label{table:correlations}
\end{table*}

A potentially important caveat to applying this result to the MW is that, as Fig.~\ref{fig:sim_vs_obs_grad} showed, our simulated gradients at $z = 0$ are likely shallower than the MW. If this discrepancy persisted across time, then the gradient of the MW was steeper than these simulations predict, which suggests a potentially earlier transition age, $\gtrsim 10.8 \Gyr$.
However, we caution that the strength of the gradient at $z = 0$ does not necessarily correlate strongly with its behavior many Gyrs ago, as we show below.

\subsection{What determines the present-day radial gradient?}
\label{subsec:gradient_correlations}

\begin{figure*}
    \centering
    \includegraphics[width = .95 \linewidth]{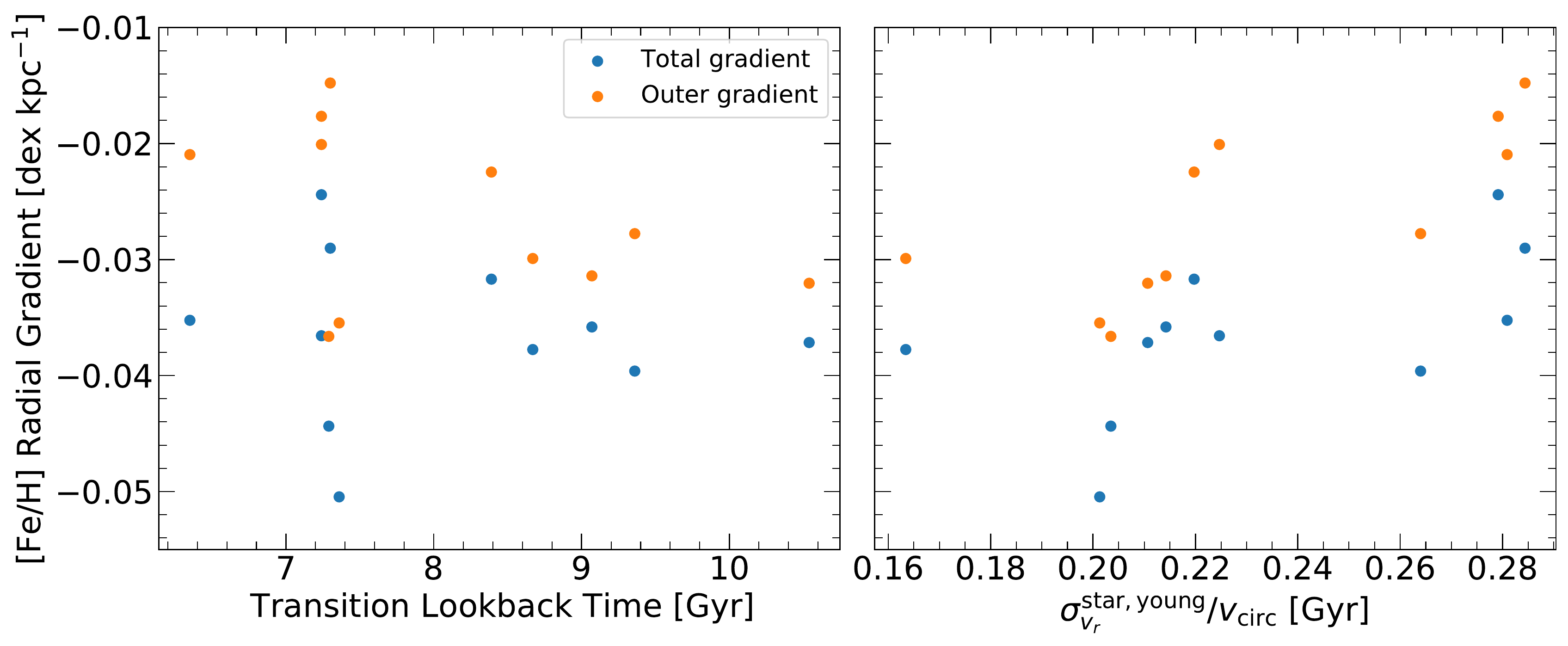}
    \vspace*{-3.5 mm}
    \caption{
    Correlation of the radial gradient in stars younger than $500 \Myr$ against various galaxy properties. The blue points show the total gradient and the orange points show the outer gradient, as discussed in Section~\ref{subsec:observation_comparison}.  \textbf{Left}: The radial gradient versus the transition time from Fig.~\ref{fig:azimuthal_scatter_vs_gradient} (right).
    While galaxies that transitioned at earlier lookback times have slightly steeper gradients at $z = 0$, on average, the correlations are not statistically significant (see Table~\ref{table:correlations}).
    As Table~\ref{table:correlations} shows, the strength of the radial gradient at $z = 0$ has little to no significant correlation with any metric of disk `settling' time, although it does show a reasonably strong correlation with overall stellar age.
    \textbf{Right}: The strongest correlation with the radial abundance gradients at $z = 0$ is with $\sigma^{\rm star, young}_{v_{r}} / v_{\rm circ}$. Galaxies that are more rotationally dominated have the stronger radial gradients in abundance, likely because of less radial mixing.}
    \label{fig:grad_vs_transition_time}
\end{figure*}

To understand better what aspect of formation history determines, or at least correlates with, the strength of the radial abundance gradient at $z = 0$, we calculate the Spearman rank correlation coefficient of the radial gradient (both total and outer) of the youngest stars in our galaxies with a variety of different metrics. Table~\ref{table:correlations} shows all metrics and associated correlations and $p$-values.

Fig.~\ref{fig:grad_vs_transition_time} shows example scatter plots for two correlation metrics. Blue points show the slope of the total gradient and the orange points show the slope of the outer gradient. The left panel shows the correlation of the \FeH{} gradients with transition lookback time (described in Section~\ref{subsec:azimuthal_v_radial}). The right panel shows the most significant correlation we find, the correlation between \FeH{} gradients and the ratio of radial velocity dispersion to circular velocity for stars younger than $500 \Myr$ ($\sigma^{\rm star, young}_{v_{r}} / v_{\rm circ}$).

We find no significant correlation between the radial gradients and the transition time from bursty to smooth star formation in these galaxies \citep[as presented in][]{Yu21}, nor with the transition times we present in Section~\ref{subsec:azimuthal_v_radial}. This indicates that the steepness of the radial abundance gradient does not depend on any metric of when the disk `settled', such as the amount of time the galaxy has experienced smooth star formation or the time since the radial gradient became the dominant source of abundance inhomogeneity.
We also find no significant correlation between the radial gradients and the size of the galaxy (measured using all or young stars), or with the strength of the gradient of the ratio of young-star to gas surface density (both measured from $R = 3 \kpc$ to $R = R^{*}_{90}$ to exclude the bulge region). Although Fig.~\ref{fig:abundance_and_mass_distributions} shows general agreement between the \textit{average} shape of $\Sigma^{\rm star, young} / \Sigma^{\rm gas}$ and the \textit{average} abundance profile across our suite, there is significant host-to-host scatter.

We also find no correlation between the radial gradient and the circularity parameter \citep[see][]{Abadi03} of young stars (age $\lesssim 500 \Myr$), defined as the ratio of the angular momentum of a star to the angular momentum with the same energy on a circular orbit.

We do find a statistically significant correlation of the radial gradient with the radial velocity dispersion of stars younger than $500 \Myr$ ($\sigma^{\rm star, young}_{v_{r}}$) and with the ratio of radial velocity dispersion to circular velocity for young stars ($\sigma^{\rm star, young}_{v_{r}} / v_{\rm circ}$). These metrics of `diskiness' at $z \sim 0$ imply that the dynamics of stars is an important factor in setting the radial gradient strength, even though the circularity parameter alone is not significant. In galaxies with significant radial velocity dispersion in young stars, there must be more radial mixing in gas (a newly formed star particles takes on the kinematic properties of its progenitor gas cell). This causes the gradients to be shallower. This implies that the degree of radial mixing (via turbulence, spiral arms, and so on) at $z \sim 0$ is more important in determining the radial gradient of abundances than the formation history.

That said, we do find one significant correlation with a formation-history metric: a negative correlation with the median stellar age of the galaxy.
This implies that galaxies whose stars formed earlier have steeper gradients at present day. This agrees with previous analysis of low mass galaxies in FIRE \citep[for example][]{Mercado21}.
But interestingly this correlation does not extend to any metric of disk settling time.
In other words, we find a correlation with when the stars formed, but not with when they formed in a settled disk.
We defer a more detailed analysis of the relationship between disk settling time and star-formation history to future work.

\subsection{Fits to functional forms}
\label{subsec:functional_forms}

\begin{table*}
\centering
\caption{
Fits for each simulation to stellar \FeH{} and \MgH{} (shown in Fig.~\ref{fig:galaxy_property_evolution}) versus age, using newly formed, in-situ stars across the galaxy. We fit the profile of each host to $\text{[X/H]} = A - Be^{t_{\rm lb} / \tau}$, where $t_{\rm lb}$ is stellar age (lookback time). The bottom row shows the fit to the mean across these 11 galaxies.
}
\begin{tabular}{l|lll|lll}
\toprule
\multicolumn{1}{c|}{\multirow{2}{*}{sim}} & 
\multicolumn{3}{c|}{{[}Fe/H{]}} & 
\multicolumn{3}{c}{{[}Mg/H{]}} \\
\multicolumn{1}{c|}{} &
\multicolumn{1}{c}{A [dex]} & 
\multicolumn{1}{c}{B [dex]} & 
\multicolumn{1}{c|}{$\tau$ [Gyr]} & 
\multicolumn{1}{c}{A [dex]} & 
\multicolumn{1}{c}{B [dex]} & 
\multicolumn{1}{c}{$\tau$ [Gyr]} \\ 
\midrule
m12m & 
$3.1 \times 10^{-1}$ & $1.0 \times 10^{-1}$ & $4.0$  & 
$5.0 \times 10^{-1}$ & $6.9 \times 10^{-2}$ & $3.7$
\\
Romulus & 
$1.5 \times 10^{-1}$ & $5.0 \times 10^{-2}$ & $3.6$ &
$3.3 \times 10^{-1}$ & $2.8 \times 10^{-2}$ & $3.2$
\\
m12b & 
$7.4 \times 10^{-2}$ & $1.1 \times 10^{-2}$ & $2.4$ & 
$2.8 \times 10^{-1}$ & $5.6 \times 10^{-3}$ & $2.2$ 
\\
m12f & 
$1.6 \times 10^{-2}$ & $1.5 \times 10^{-2}$ & $2.6$ & 
$2.4 \times 10^{-1}$ & $9.3 \times 10^{-3}$ & $2.4$
\\
Thelma & 
$3.8 \times 10^{-1}$ & $2.8 \times 10^{-1}$ & $6.0$ & 
$5.6 \times 10^{-1}$ & $2.2 \times 10^{-1}$ & $5.5$ 
\\
Romeo & 
$6.8 \times 10^{-2}$ & $1.4 \times 10^{-2}$ & $2.7$ & 
$2.5 \times 10^{-1}$ & $5.2 \times 10^{-3}$ & $2.3$ 
\\
m12i & 
$1.2 \times 10^{-1}$ & $5.8 \times 10^{-2}$ & $3.6$ & 
$3.2 \times 10^{-1}$ & $4.0 \times 10^{-2}$ & $3.3$
\\
m12c & 
$2.5 \times 10^{-1}$ & $1.5 \times 10^{-1}$ & $5.1$ & 
$4.3 \times 10^{-1}$ & $1.1 \times 10^{-1}$ & $4.6$ 
\\
Remus & 
$8.1 \times 10^{-2}$ & $4.8 \times 10^{-2}$ & $3.8$ & 
$2.4 \times 10^{-1}$ & $2.3 \times 10^{-2}$ & $3.2$
\\
Juliet & 
$9.9 \times 10^{-3}$ & $1.8 \times 10^{-2}$ & $2.9$ & 
$1.9 \times 10^{-1}$ & $9.3 \times 10^{-3}$ & $2.5$
\\
Louise & 
$5.2 \times 10^{-2}$ & $8.8 \times 10^{-2}$ & $4.4$ & 
$2.0 \times 10^{-1}$ & $5.1 \times 10^{-2}$ & $3.8$
\\
\midrule
Mean & 
$1.1 \times 10^{-1}$ & $5.4 \times 10^{-2}$ & $3.6$ & 
$3.0 \times 10^{-1}$ & $3.3 \times 10^{-2}$ & $3.2$
\\
\bottomrule
\end{tabular}
\label{table:xonh_evolution_fits}
\end{table*}

\begin{table*}
\centering
\caption{
Fits to the radial gradient ($\Delta \text{[X/H]}_{R^{*}_{90}} / R^{*}_{90}$) of stars at formation versus lookback time for each simulation, for both \FeH{} and \MgH.
We fit to a second-order polynomial: $\Delta \text{[X/H]}_{R^{*}_{90}} / {R^{*}_{90}} = A t^2_{\rm lb} + Bt_{\rm lb} + C$.
The bottom shows the fit to the mean trend across all 11 galaxies.
}
\begin{tabular}{l|rrr|rrr}
\toprule
\multicolumn{1}{c|}{\multirow{2}{*}{sim}} & 
\multicolumn{3}{c|}{{[}Fe/H{]}} & 
\multicolumn{3}{c}{{[}Mg/H{]}} \\
\multicolumn{1}{c|}{} &
\multicolumn{1}{c}{A $\left[ \frac{\dex}{\kpc \Gyr^2} \right]$} & 
\multicolumn{1}{c}{B $\left[ \frac{\dex}{\kpc \Gyr} \right]$} & 
\multicolumn{1}{c|}{C $\left[ \frac{\dex}{\kpc} \right]$} & 
\multicolumn{1}{c}{A $\left[ \frac{\dex}{\kpc \Gyr^2} \right]$} & 
\multicolumn{1}{c}{B $\left[ \frac{\dex}{\kpc \Gyr} \right]$} & 
\multicolumn{1}{c}{C $\left[ \frac{\dex}{\kpc} \right]$} \\ 
\midrule
m12m & 
$-4.5 \times 10^{-5}$ & $3.1 \times 10^{-3}$ & $-3.5 \times 10^{-2}$ & 
$-4.5\times 10^{-5}$ & $1.9 \times 10^{-3}$ & $-3.0 \times 10^{-2}$ 
\\
Romulus & 
$-1.2 \times 10^{-4}$ & $2.6 \times 10^{-3}$ & $-3.4 \times 10^{-2}$ & 
$-1.4 \times 10^{-7}$ & $1.1 \times 10^{-3}$ & $-2.6 \times 10^{-2}$ 
\\
m12b & 
$1.3 \times 10^{-3}$ & $-1.2 \times 10^{-2}$ & $-3.5 \times 10^{-2}$ & 
$1.2 \times 10^{-3}$ & $-1.2 \times 10^{-2}$ & $-2.6 \times 10^{-2}$ 
\\
m12f & 
$4.1 \times 10^{-4}$ & $-2.3 \times 10^{-3}$ & $-3.0 \times 10^{-2}$ & 
$3.6 \times 10^{-4}$ & $-2.1 \times 10^{-3}$ & $-2.4 \times 10^{-2}$ 
\\
Thelma & 
$3.8 \times 10^{-4}$ & $-8.7 \times 10^{-4}$ & $-2.4 \times 10^{-2}$ & 
$6.1 \times 10^{-4}$ & $-3.3 \times 10^{-3}$ & $-1.8 \times 10^{-2}$ 
\\
Romeo & 
$3.3 \times 10^{-4}$ & $-1.1 \times 10^{-3}$ & $-3.8 \times 10^{-2}$ & 
$4.0 \times 10^{-4}$ & $-2.5 \times 10^{-3}$ & $-2.7 \times 10^{-2}$ 
\\
m12i & 
$3.6 \times 10^{-4}$ & $9.7 \times 10^{-6}$ & $-3.6 \times 10^{-2}$ & 
$3.7 \times 10^{-4}$ & $-7.7 \times 10^{-4}$ & $-2.8 \times 10^{-2}$ 
\\
m12c & 
$1.4 \times 10^{-4}$ & $2.1 \times 10^{-3}$ & $-3.7 \times 10^{-2}$ & 
$2.2 \times 10^{-4}$ & $7.8 \times 10^{-4}$ & $-3.0 \times 10^{-2}$ 
\\
Remus & 
$3.2 \times 10^{-4}$ & $-1.3 \times 10^{-3}$ & $-3.6 \times 10^{-2}$ & 
$3.1 \times 10^{-4}$ & $-1.7 \times 10^{-3}$ & $-2.8 \times 10^{-2}$ 
\\
Juliet & 
$-1.9 \times 10^{-4}$ & $6.3 \times 10^{-3}$ & $-5.5 \times 10^{-2}$ & 
$-1.1 \times 10^{-4}$ & $4.6 \times 10^{-3}$ & $-4.3 \times 10^{-2}$ 
\\
Louise & 
$3.8 \times 10^{-4}$ & $2.9 \times 10^{-4}$ & $-4.3 \times 10^{-2}$ & 
$4.9 \times 10^{-4}$ & $-9.6 \times 10^{-4}$ & $-3.6 \times 10^{-2}$ 
\\ 
\midrule
Mean & 
$3.0 \times 10^{-4}$ & $-3.0 \times 10^{-4}$ & $-3.7 \times 10^{-2}$ & 
$3.6 \times 10^{-4}$ & $-1.3 \times 10^{-3}$ & $-2.9 \times 10^{-2}$ 
\\ 
\bottomrule
\end{tabular}
\label{table:radial_gradient_fits}
\end{table*}

\begin{table*}
\centering
\caption{
Fits to the total ($360^{\circ}$) azimuthal scatter of stars at formation, averaged across all radii, versus lookback time for each simulation. We fit a second-order polynomial $\sigma^{360^{\circ}}_{\text{[X/H]}}$ $ = A t_{\rm lb}^2 + B t_{\rm lb} + C$.
The bottom shows the fit to the mean trend across all 11 galaxies.
}
\begin{tabular}{l|rrr|rrr}
\toprule
\multicolumn{1}{c|}{\multirow{2}{*}{sim}} & 
\multicolumn{3}{c|}{{[}Fe/H{]}} & 
\multicolumn{3}{c}{{[}Mg/H{]}} \\
\multicolumn{1}{c|}{} &
\multicolumn{1}{c}{A $\left[ \frac{\dex}{\Gyr^2} \right]$} & 
\multicolumn{1}{c}{B $\left[ \frac{\dex}{\Gyr} \right]$} & 
\multicolumn{1}{c|}{C $\left[ \dex \right]$} & 
\multicolumn{1}{c}{A $\left[ \frac{\dex}{\Gyr^2} \right]$} & 
\multicolumn{1}{c}{B $\left[ \frac{\dex}{\Gyr} \right]$} & 
\multicolumn{1}{c}{C $\left[ \dex \right]$} \\ 
\midrule
m12m    & 
$1.3 \times 10^{-3}$  & $-1.1 \times 10^{-3}$  & $3.3 \times 10^{-2}$ & 
$1.6 \times 10^{-3}$  & $-5.3 \times 10^{-3}$  & $3.2 \times 10^{-2}$
\\
Romulus & 
$9.8 \times 10^{-4}$ & $-6.0 \times 10^{-3}$ & $4.4 \times 10^{-2}$ &
$1.0 \times 10^{-3}$ & $-5.3 \times 10^{-3}$  & $3.5 \times 10^{-2}$
\\
m12b    & 
$6.7 \times 10^{-4}$ & $-3.0 \times 10^{-3}$  & $4.6 \times 10^{-2}$ &
$7.3 \times 10^{-4}$ & $-2.5 \times 10^{-3}$  & $3.8 \times 10^{-2}$
\\
m12f    & 
$1.4 \times 10^{-3}$  & $-3.7 \times 10^{-3}$ & $7.2 \times 10^{-2}$ &
$1.4 \times 10^{-3}$  & $-2.7 \times 10^{-3}$  & $6.0 \times 10^{-2}$
\\
Thelma  & 
$1.2 \times 10^{-3}$ & $-2.5 \times 10^{-4}$ & $3.3 \times 10^{-2}$ &
$1.2 \times 10^{-3}$ & $-3.1 \times 10^{-4}$  & $2.8 \times 10^{-2}$
\\
Romeo   & 
$1.7 \times 10^{-3}$ & $-8.5 \times 10^{-3}$ & $6.6 \times 10^{-2}$ &
$1.7 \times 10^{-3}$ & $-7.5 \times 10^{-3}$  & $5.1 \times 10^{-2}$
\\
m12i    & 
$1.4 \times 10^{-3}$ & $-1.0 \times 10^{-2}$ & $4.8 \times 10^{-2}$ &
$1.4 \times 10^{-3}$ & $-8.9 \times 10^{-3}$  & $4.1 \times 10^{-2}$
\\
m12c    & 
$7.6 \times 10^{-4}$ & $-3.6 \times 10^{-3}$ & $4.4 \times 10^{-2}$ &
$9.0 \times 10^{-4}$ & $-4.2 \times 10^{-3}$  & $4.0 \times 10^{-2}$
\\
Remus   & 
$5.3 \times 10^{-4}$ & $-1.6 \times 10^{-3}$ & $4.0 \times 10^{-2}$  &
$5.7 \times 10^{-4}$ & $-9.3 \times 10^{-4}$  & $3.2 \times 10^{-2}$
\\
Juliet  & 
$1.6 \times 10^{-3}$ & $-1.2 \times 10^{-2}$ & $9.4 \times 10^{-2}$ &
$1.4 \times 10^{-3}$ & $-8.3 \times 10^{-3}$  & $7.7 \times 10^{-2}$
\\
Louise  & 
$1.5 \times 10^{-3}$ & $-5.2 \times 10^{-3}$  & $4.9 \times 10^{-2}$ &
$1.7 \times 10^{-3}$ & $-5.7 \times 10^{-3}$  & $4.2 \times 10^{-2}$
\\
\midrule
Mean    & 
$1.2 \times 10^{-3}$  & $-4.6 \times 10^{-3}$ & $5.2 \times 10^{-2}$ &
$1.2 \times 10^{-3}$  & $-4.1 \times 10^{-3}$  & $4.3 \times 10^{-2}$
\\ 
\bottomrule
\end{tabular}
\label{table:azimuthal_scatter_fits}
\end{table*}

For stars at the time of their formation, we quantify the evolution of the galaxy-wide abundance across time (Fig.~\ref{fig:galaxy_property_evolution} middle), the overall radial gradient across time (Fig.~\ref{fig:stellar_radial_grad_evolution} top), and the $360^{\circ}$ azimuthal scatter across time (Fig.~\ref{fig:azimuthal_scatter_vs_gradient} left). We fit each galaxy independently, as well as the mean trends across all 11 galaxies.
We tested fitting as a function of redshift, expansion scale factor, and stellar age.  We found fitting as a function of stellar age to provide the best fits.

Table~\ref{table:xonh_evolution_fits} shows the best fit to the average \FeH{} and \MgH{} of all stars at formation within the galaxy across time, that is, the fit to Fig.~\ref{fig:galaxy_property_evolution} (middle). We fit to the functional form:
\begin{equation}
\label{eq:metal_evolution}
    \mathrm{[X/H]} = A - Be^{t_{\rm lb} / \tau}
\end{equation}
We determine the best fit coefficients using the optimize.curve\_fit function in SciPy. We tested second-order polynomial fits as well as simple exponential fits, but we find better agreement with the functional form in Eq.~\ref{eq:metal_evolution}.

Table~\ref{table:radial_gradient_fits} shows the best fit to the time evolution of the overall radial gradient, $\Delta \text{[X/H]}_{R^{*}_{90}} / R^{*}_{90}$. We fit a second-order polynomial:
\begin{equation}
    \Delta \text{[X/H]}_{R^{*}_{90}}/R^{*}_{90} = At^2_{\rm lb}+Bt_{\rm lb}+C
\end{equation}
We determine the best fit coefficients using the polyfit function from NumPy. We select a second-order fit, because the radial gradient evolution of some galaxies is too complicated to be captured by a linear fit, and a 3rd-order polynomial suffers from overfitting.

Table~\ref{table:azimuthal_scatter_fits} shows the best fits to the evolution of the $360^{\circ}$ azimuthal scatter of \FeH. We average the azimuthal scatter over $5$ radii (2, 4, 6, 8 , and 10 kpc).  We fit a 2nd order polynomial using the polyfit function in NumPy, as with the radial gradient evolution:
\begin{equation}
\sigma^{360^{\circ}}_{[\text{X/H]}} = At^2_{\rm lb} + Bt_{\rm lb} + C
\end{equation}
We tested an exponential fit too, but we find better agreement for a second-order polynomial.

\subsection{Comparison to previous results for all gas}

\citetalias{Bellardini21} examined the evolution of elemental abundance variations for all gas (not just star-forming gas) in the same MW-mass FIRE-2 simulations. Here, we expand on those results by specifically examining newly formed stars. We generally expect the trends for newly formed stars to match those of the gas, and while we find overall qualitatively similar results, we briefly summarize key quantitative differences and similarities.

For radial gradients, we find generally steeper radial gradients in newly formed stars relative to all gas.
For vertical gradients, both all gas and newly formed stars exhibit little to no variation with height.
For azimuthal scatter, newly formed stars are weaker than that of all gas. This indicates a higher degree of homogeneity for star-forming gas. Additionally, Fig.~\ref{fig:stellar_azimuthal_scatter_evolution} shows that the azimuthal scatter of abundances for newly formed stars is independent of radius. This contrasts the results of \citetalias{Bellardini21}, who found that the azimuthal scatter of all gas increases with increasing radius.
Appendix~\ref{appendix:stars_vs_gas} directly compares the evolution of azimuthal scatter and radial change in abundance for newly formed stars versus all gas, following the methods of \citetalias{Bellardini21}.

Following the analysis of \citetalias{Bellardini21} (their Section~3.5), Appendix.~\ref{appendix:measurable_homogeneity} shows the radial scale at which azimuthal abundance variations are subdominant to the radial variations in abundance ($\Delta R_{\rm equality}$) and the radial scale at which radial variations are measurable homogeneous ($\Delta R_{\rm homogeneous}$. $\Delta R_{\rm equality}$ and $\Delta R_{\rm homogeneous}$ for newly formed stars are smaller than for all gas, which reflects the steeper radial gradients and smaller azimuthal scatter in newly formed stars.

\section{Summary and Discussion}
\label{sec:sum_and_disc}

\subsection{Summary}
\label{subsec:sum}

We used a suite of $11$ MW/M31-mass cosmological zoom-in simulations, run with FIRE-2 physics, to explore the 3-D spatial variations of elemental abundances of stars \textit{at birth} (within $< 50 \Myr$ of their formation) across these galaxies' formation histories.
(In future work we will examine similar trends for stellar populations at $z = 0$ as a function of their age.)
We measured properties of newly formed stars as a function of lookback time going back $\approx 12 \Gyr$, in part to test and guide approaches to chemical tagging.
We also fit functional forms to these trends, to use in models of elemental evolution.
Our main results are:

\begin{itemize}
    \item \textit{Galaxy stellar abundances}: enrich relatively quickly: 5 of our 11 galaxies reached $\FeH \approx -0.5$ at lookback times of $9 - 10 \Gyr$ ago. LG-like galaxies enriched in metals faster than isolated galaxies, following their more rapid stellar mass assembly (see Fig.~\ref{fig:galaxy_property_evolution} as well as \citealt{Santistevan20}).

    \item \textit{Galaxy stellar size}: $R^{*}_{90}$ for both young stars and all stars increased over time. $R^{*}_{90}$ for all stars was comparable to that of young stars $\gtrsim 7.5 \Gyr$ ago. However, after this, $R^{*}_{90}$  of young stars is systematically larger than that of all stars, reflecting inside-out radial growth. For $t_{\rm lb} \lesssim 8 \Gyr$, using either all stars or young stars, $R^{*}_{90}$ is larger for galaxies in LG-like environments than those that are isolated ($16.2 \kpc$ versus $13.1 \kpc$ for young stars and $11.1 \kpc$ versus $10.2 \kpc$ for all stars at $z = 0$; see also \citealt{Garrison-Kimmel18}).

    \item \textit{Galaxy-wide scatter in abundances}: reached a minimum of $\approx 0.09 \dex$ for in-situ stars forming $\approx 7 \Gyr$ ago. This reflects a competition between a reduction of the scatter as azimuthal variations decreased over time and an increase in the galaxy-wide scatter as the radial gradient became stronger.

    \item \textit{Vertical gradients}: at formation are negligible in nearly all regimes.
    The change in abundance is on average less than $0.02 \dex$ over $1 \kpc$ for both \FeH{} and \MgH. Thus, vertical abundance variations provide minimal discriminating power for chemical tagging.
    The one exception that we find is for stars in the inner bulge region, $R \lesssim 4 \kpc$, at $z \approx 0$.

    \item \textit{Radial gradients}: of newly formed stars were flat (magnitude $\lesssim 0.01 \dpk$) at lookback times $\gtrsim 9.5 \Gyr$ but became progressively steeper with time, reaching $-0.037 \dpk$ for \FeH{} ($-0.030 \dpk$ for \MgH) at $z = 0$.
    \FeH{} gradients for newly formed stars at $z = 0$ are shallower than in the MW measured over similar radial ranges.  However, \citetalias{Bellardini21} showed that gas-phase radial abundance gradients in our simulations are as steep or steeper than those observed in external MW-mass galaxies.  Our galaxies are well fit by two-component radial gradients that are steeper in the inner galaxy, which reflects a steeper stellar surface density profile in the inner galaxy, though not necessarily with the bulge region.

    \item \textit{Azimuthal scatter}: of young stars systematically decreases over time from $\lesssim 0.18 \dex$ $11.6 \Gyr$ ago to $\lesssim 0.043 \dex$ today for \FeH{} and \MgH. Azimuthal scatter shows minimal dependence on azimuthal bin size, so small-scale variations dominate over larger-scale variations.
    Even at scales $\lesssim 1 \kpc$ we measure \FeH{} azimuthal scatter in the solar cylinder of $\approx 0.043 \dex$ at $z = 0$ and $\approx 0.16 \dex$ for stars that formed at $t_{\rm lb} = 11.6 \Gyr$. Importantly, our analysis does \textit{not} center on individual star-forming regions, but rather, random patches in the galaxy, so this is not a statement about the internal homogeneity of individual star-forming regions and star clusters.

    \item \textit{Azimuthal versus radial variations}: Similar to our analysis of gas in \citetalias{Bellardini21}, our simulated galaxies transitioned from being dominated by azimuthal scatter to being dominated by radial variations at lookback times of $\approx 8 \Gyr$ ago.
    Thus, azimuthal variations were the primary source of galaxy-wide scatter in abundance at early times, and they are of secondary importance (though not negligible) for stars that formed $\lesssim 8 \Gyr$ ago.

    \item \textit{Correlations with present-day radial gradient}: We tested the correlation between the strength of the radial gradient at $z = 0$ and a variety of metrics of formation history. The most statistically significant correlation is the ratio of the radial velocity dispersion of young stars to their circular velocity. So, the degree of radial mixing in galaxies is likely set by the strength of the ratio of radial velocity dispersion to circular velocity.  Additionally, we find a lack of correlation with the transition lookback time (see Section~\ref{subsec:azimuthal_v_radial}) which implies disk settling time is not responsible for setting present-day abundance gradients.

    \item \textit{Fit to functional forms}: We fit the evolution of overall normalization, the radial gradient, and the azimuthal scatter of the abundances of stars at formation versus lookback time (stellar age) in Section~\ref{subsec:functional_forms}.
\end{itemize}

\subsection{Limitations and caveats}
\label{subsec:caveats}

The simulations analyzed in this work implement the FIRE-2 physics, discussed in Section~\ref{subsec:sims}.  \citet{Hopkins18} presents the physics in detail and a variety of tests showing their robustness robustness.  However there are still limitations inherent to our analysis and the physics implemented in FIRE-2.

We analyze only $11$ galaxies, so our results are limited by our sample size. Furthermore, because we chose these galaxies to be near the mass of the MW, they necessarily encompass a narrow range of stellar and halo masses  (see Table~\ref{table:general_host_properties}).

Also, as (\citet{Sanderson20} and McCluskey et al. in prep) show, the velocity dispersion of stars in these simulations is dynamically hotter than observed in the MW (though they are more similar to M31). This could play a role in the shallower radial gradients in our simulations (see Section~\ref{subsec:observation_comparison}), given the strong correlation between stellar velocity dispersion and strength of the radial gradient (see Section~\ref{subsec:gradient_correlations}).

In addition, there are limitations in our current physics implementations. The simulations do not include a self-consist treatment of cosmic rays or magnetohydrodynamics and anisotropic thermal conduction and viscosity in gas \citep[see][]{Hopkins18}. FIRE-2 treats all core-collapse supernovae as having identical IMF-averaged yields, but different mass progenitors will have different yields \citep[see][]{Muley21}.  Future FIRE-3 simulations will appropriately mass sample rates and yields of different mass core-collapse supernovae \citep[][]{Hopkins22}.  Additionally, \citet{Gandhi22} showed that the default implementation of type Ia supernovae rates in FIRE-2 may be underestimated, leading to an underproduction of \FeH{}.

Finally, these FIRE-2 simulations do not include any treatment of AGN from supermassive black holes, which may bias the dynamics and star-formation rates, particularly in the inner few kpc.
However, recent implementations in FIRE \citep[][]{Wellons22} will allow us to explore their effects in future work.

\subsection{Discussion}
\label{subsec:disc}

Our analysis extends the work in \citetalias{Bellardini21}, in which we examined the homogeneity of gas, as an initial guide for the homogeneity of newly formed stars. The primary goal of this paper is to quantify the elemental abundance homogeneity of newly formed stars as a function of lookback time and to provide functional forms to its evolution such that chemical-tagging models have better descriptors of the initial degree of homogeneity with which stars form in a cosmological context. This will allow for more accurate galactic elemental evolution models and provide more realistic expectations for chemical tagging models.

We examined \FeH, \MgH, and \MgFe{}. We look at Fe primarily as a representative element of type Ia supernovae and Mg as a representative $\alpha$ element, that is, primarily sourced via core collapse supernovae. Because our analysis is limited to only $2$ elements, we do not asses if analysis of more abundances will provide more discriminating power for chemical tagging. However, the work of \citet{Ting21} suggests including at least $7-8$ elements when doing galactic archaeology, \citet{Casamiquela21} concluded that the larger the abundance space the better for chemical tagging, and \citet{Ratcliffe21} found that stellar clusters are better identified when using $15$ abundances rather than $2$. Our FIRE-2 simulations track $9$ metals. Given the expected correlation between elements primarily sourced via the same enrichment channels, we defer a more detailed analysis of all $9$ elements to future work using the FIRE-3 simulations \citep[][]{Hopkins22}, which implement a tracer-element approach for varying stellar yields in post-processing (Wetzel et al. in prep).

As we showed in Section~\ref{subsec:general_properties}, the average \FeH{} of stars in our galaxies as well as the average stellar mass of our galaxies are similar to those of the MW. Additionally, we measured the scatter in \FeH{} as a function of stellar age and found it decreases with decreasing stellar age for stars older than $\approx 7 \Gyr$ and then increases with decreasing stellar age.  This is discrepant with previous analysis of stellar metallicity distributions in the MW that find the scatter in \FeH{} continually decreases with decreasing stellar age \citep{Casagrande11, Miglio21}. However, we do not match the selection function of these surveys, nor do we divide our stars into similar age bins.

However, an important caveat to this work is that our simulations are not designed to recreate the history of the MW; they instead provide a cosmologically representative range of histories of galaxies that are similar to the MW at $z = 0$. \citet{Boardman20a} suggest that, because the MW has a particularly small disc scale length relative to similar mass galaxies, it is important to factor in disc scale length when selecting MW analogs. However, in \citetalias{Bellardini21} we tested scaling the gas-phase radial abundance gradients in our MW-mass galaxies and found the most self-similarity when measuring gradients in physical units.

We explore the effect of varying the age range of stars used to measure the radial gradients in the simulations, because the gradients are systematically shallower than observations of the MW, although our gradients are steeper than those measured in nearby MW-mass galaxies. This in principle could account for uncertainties in the ages of observed star clusters. However, including older populations of stars leads to increasingly shallower gradients, resulting in greater discrepancies with observations. Some observations \citep[for example][]{Carrera11, Cunha16, Netopil16, Donor20, SP21} find that the MW's galactic radial gradient in abundance as determined by older star clusters is steeper than the gradient as determined by younger star clusters, but our results imply the opposite. Our results agree with \OH{} measurements from planetary nebulae in M81 \citep[for example][]{Stanghellini10b, Stanghellini14} and in the MW \citep{Stanghellini18}. However, we do not measure the radial positions of stars at $z = 0$ as a function of age, although we plan to pursue this in future work. For our comparison we examine the effect of widening the age cutoff we use to identify `young' stars at $z = 0$ from $< 0.5 \Gyr$ to $< 10 \Gyr$. The shallower radial gradients when including older stars likely results from radial mixing \citep[for example][]{SB09a, Loebman11, Quillen18}.

The steepening of the radial gradients in our simulations disagrees with some observations which indicate either a flattening of radial abundance gradients \citep[for example][]{Frinchaboy13, Netopil16, Spina17} or a steepening and then flattening \citep[for example][]{Xiang15, Xiang17, Anders17} but agrees with others \citep[for example][]{Stanghellini10a, Stanghellini18}. However, present-day measurements of stellar positions do not necessarily represent the formation locations of stars, so these results are not directly comparable to our work. A better point of comparison is either spatially resolved observations of stars in MW progenitor analogues, or spatially resolved gas-phase abundance observations \citepalias[as discussed in][]{Bellardini21}, assuming the stars form with abundances representative of the gas. High-redshift observations of gas-phase abundance gradients generally show flat radial profiles \citep[for example][]{Wuyts16, Patricio19, Curti20}, but some show strong negative radial gradients \citep[for example][]{Carton18, Wang20}.  At high redshifts, galaxies do not sustain gas -phase abundance gradients \citepalias{Bellardini21}. Once disk settling occurs, gas no longer mixes radially as efficiencly in the disk, so it can take on different abundances at different radii, imprinted on newly formed stars.

In addition to our fiducial measure of the total radial gradient, we fit a two-component piecewise linear function to the abundance profile.  We generally find that the inner region of the disk is steeper than the outer region, in contrast to some observations \citep[for example][]{Hayden14, Maciel19}, but in agreement with other observations \citep[for example][]{Netopil16, Reddy20}. Exact comparisons with observations are difficult, because observations typically measure the radial gradient for all stars, rather than measuring for mono-age stellar populations. Also, uncertainties in the inferred ages and locations of stars can influence the interpreted gradients.

One of the key results of this paper is the quantification of the evolution of the azimuthal scatter in abundance of newly formed stars. The azimuthal scatter does not directly track the scatter in the gas in \citetalias{Bellardini21}. The scatter is systematically smaller for stars than for all gas. This discrepancy is larger at larger lookback times and larger radii (see Fig.~\ref{fig:azimuthal_v_grad_gas_v_stars}) which results from stars at large lookback times and large radii preferentially forming from gas occupying slightly higher metallicity. This tests a common assumption in chemical-tagging models, that stars have abundances that primarily depend just on their birth radii and not on azimuthal position \citep[for example][]{Frankel18, Frankel20}.

Azimuthal scatter of stellar abundances is important to quantify for chemical tagging, but has not yet been well characterized by observations. Observations of cepheids in the MW by \citet{Luck11a} indicate no significant azimuthal dependence, which is consistent with the small azimuthal scatter we measure for the youngest stars in Fig.~\ref{fig:young_stellar_azimuthal_scatter}. However, our results also indicate much smaller azimuthal scatter than that observed in the MW by \citet{Kovtyukh22}.  Measuring cepheids at $R = 7 - 9 \kpc$ \citet{Kovtyukh22} find \FeH{} varies by up to $0.2 \dex$, much larger than our measured $0.05 \dex$ scatter.

Additionally, in this paper we quantify a transition lookback time, in the same way as \citetalias{Bellardini21}. This transition lookback time sets the timescale over which elemental evolution models reasonably can assume azimuthal homogeneity \citep[for example][]{Minchev18, Molla19a, Frankel20}. Notably, the lookback time for newly formed stars is $\sim 8 \Gyr$, $\approx 1 \Gyr$ earlier than the lookback time derived for the gas in \citetalias{Bellardini21}. This is primarily because the azimuthal scatter in stars at large lookback times were smaller than that of all gas. This may reflect the disk-wide scatter in all gas being slightly smaller than the disk-wide scatter in star-forming gas \citepalias{Bellardini21}.

Complementary to our analysis is that of \citet{Yu21}, who measured the transition epoch from `bursty' to `steady' star formation and disk settling in the same simulations. Similar to the analysis presented in \citetalias{Bellardini21}, we find that our transition times are consistently earlier than the transition times in \citet{Yu21}, by $\sim 3.2 \Gyr$ on average. However, our transition lookback times are moderately correlated with those in \citet{Yu21} (Pearson correlation coefficient $ r \approx 0.63$). Thus, the onset of a strong radial gradient in abundance precedes but correlates with the onset of `steady' star formation in these simulations.

Perhaps most important to this analysis is our characterization of the evolution of elemental abundances, radial gradients, and azimuthal scatter across our suite of MW-mass simulations (see Section~\ref{subsec:functional_forms}). We provide simple functional forms which encapsulate the evolution of abundance distributions of MW-mass galaxies across time. These functional forms and fits are \textit{crucial} to the future of elemental-evolution modeling and accurate chemical tagging, because weak chemical tagging is infeasible without an accurate picture of the birth conditions of stars.


\section*{Acknowledgements}

We thank the anonymous reviewer for the detailed comments which we feel have significantly improved the quality and clarity of the article.

We performed this work using the \textsc{GizmoAnalysis} package \citep[][]{GIZMO20}, the Astropy package \citep[][]{astropy:2013, astropy:2018}, as well as libraries from Numpy \citep[][]{Numpy20}, SciPy \citep[][]{SciPy20}, and Matplotlib \citep[][]{Matplotlib07}.

MB and AW received support from: the NSF via CAREER award AST-2045928 and grant AST-2107772; NASA ATP grants 80NSSC18K1097 and 80NSSC20K0513; HST grants AR-15809, GO-15902, GO-16273 from STScI; a Scialog Award from the Heising-Simons Foundation; and a Hellman Fellowship.
We performed this work in part at the Aspen Center for Physics, supported by NSF grant PHY-1607611.
We ran simulations using: XSEDE, supported by NSF grant ACI-1548562; Blue Waters, supported by the NSF; Frontera allocations AST21010 and AST20016, supported by the NSF and TACC; Pleiades, via the NASA HEC program through the NAS Division at Ames Research Center.

\section*{Data Availability}

The data in these figures are available at \url{https://mbellardini.github.io/}.
The FIRE-2 simulations are publicly available \citep{Wetzel2022} at \url{http://flathub.flatironinstitute.org/fire}.
Additional FIRE simulation data is available at \url{https://fire.northwestern.edu/data}.
A public version of the GIZMO code is available at \url{http://www.tapir.caltech.edu/~phopkins/Site/GIZMO.html}.



\bibliographystyle{mnras}
\bibliography{3D_stellar_abundances}



\appendix
\section{Shapes of abundance profiles}
\label{appendix:radial_profile_shapes}

\begin{figure}
	\includegraphics[width = .91 \columnwidth]{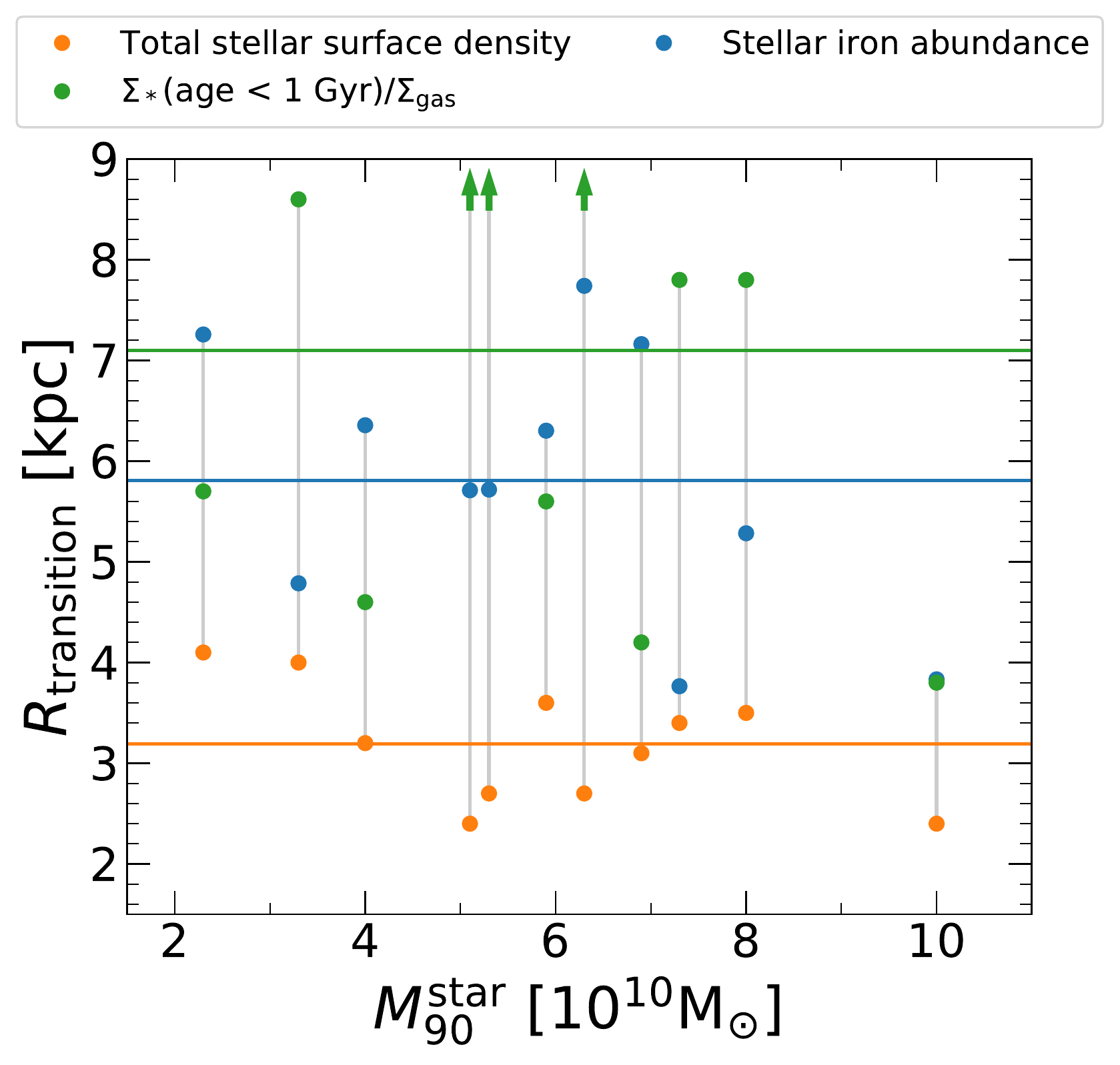}
	\vspace*{-3.5 mm}
    \caption{
    Various transition (break) radii ($R_{\rm transition}$) for each galaxy as a function of stellar mass. The orange points show $R_{\rm transition}$ of the stellar surface density profile, caused by the steeper bulge-like region, as defined in \citetalias{Bellardini21}. The blue points show $R_{\rm transition}$ from a 2-component linear fit to the \FeH{} profiles, as in Fig.~\ref{fig:current_stellar_radial_profiles}. The green points show $R_{\rm transition}$ from a 2-component linear fit (measured from $R = 3 - 12 \kpc$) to ratio of the stellar surface density to gas surface density for stars younger than $1 \Gyr$ old. The horizontal lines show the mean of each $R_{\rm transition}$.  $R_{\rm transition}$ as defined by the stellar iron abundance is always larger than $R_{\rm transition}$ as defined by the surface density of all stars. We expect a correlation between $R_{\rm transition}$ for the stellar surface density and $R_{\rm transition}$ for the stellar \FeH{} because the \FeH{} abundance of the youngest stars is a result of the supernovae from previous stellar generations.
    }
    \label{fig:break_radius_v_mass}
\end{figure}

We investigate the cause of the break in the radial abundance profiles (see Fig.~\ref{fig:current_stellar_radial_profiles}) by exploring the profiles of surface density. We fit a two-component linear profile to the log of the young stellar to gas surface ratio in Fig.~\ref{fig:abundance_and_mass_distributions} (bottom) and plot the transition radii in Fig.~\ref{fig:break_radius_v_mass} in green. In principal, the ratio of surface density of stars to gas should be approximately proportional to the abundance profile, in the limit of local metal deposition.

The blue points in Fig.~\ref{fig:break_radius_v_mass} show the transition from the steep inner radial abundance profile to the flatter outer abundance profile for each galaxy as a function of $M^{\rm star}_{90}$ (see Table~\ref{table:general_host_properties} for masses), and the horizontal line shows the average of $5.8 \kpc$ across all 11 galaxies. The orange points show the radius at which the overall stellar surface density transitions from being dominated by a s\'{e}rsic profile to being dominated by an exponential profile, which we presented in \citetalias{Bellardini21}, and which reflects the transition to a bulge-like component in each galaxy.  We found this by simultaneously fitting an exponential plus s\'{e}rseic profile to the overall stellar surface density at $z = 0$, where we fixed the s\'{e}rsic index at $n = 1.3$.

Fig.~\ref{fig:break_radius_v_mass} shows no clear stellar mass dependence to these transition radii.
The transition radius in \FeH{} for young stars is always larger than the transition radius in the surface density, typically by $\sim 2.6 \kpc$. Additionally, the transition radius of the ratio of stellar to gas surface density is, on average, larger than transition radius of the abundance profile by $\sim 1.3 \kpc$.

We thus conclude that the transition radius in the abundance profile for young stars in our simulation is not related to the onset of a bulge-like component.
Instead, it coincides better with the transition radius in the stellar to gas ratio of surface densities, and its shape (break) is at least partially set by the shape of that ratio.
However, we do not find perfect agreement.
More work is needed to understand the full shape of the abundance gradient in the context of complex metal injection, mixing, outflows, and stellar redistribution.

\section{Stars at formation versus gas}
\label{appendix:stars_vs_gas}

\begin{figure}
	\includegraphics[width = \columnwidth]{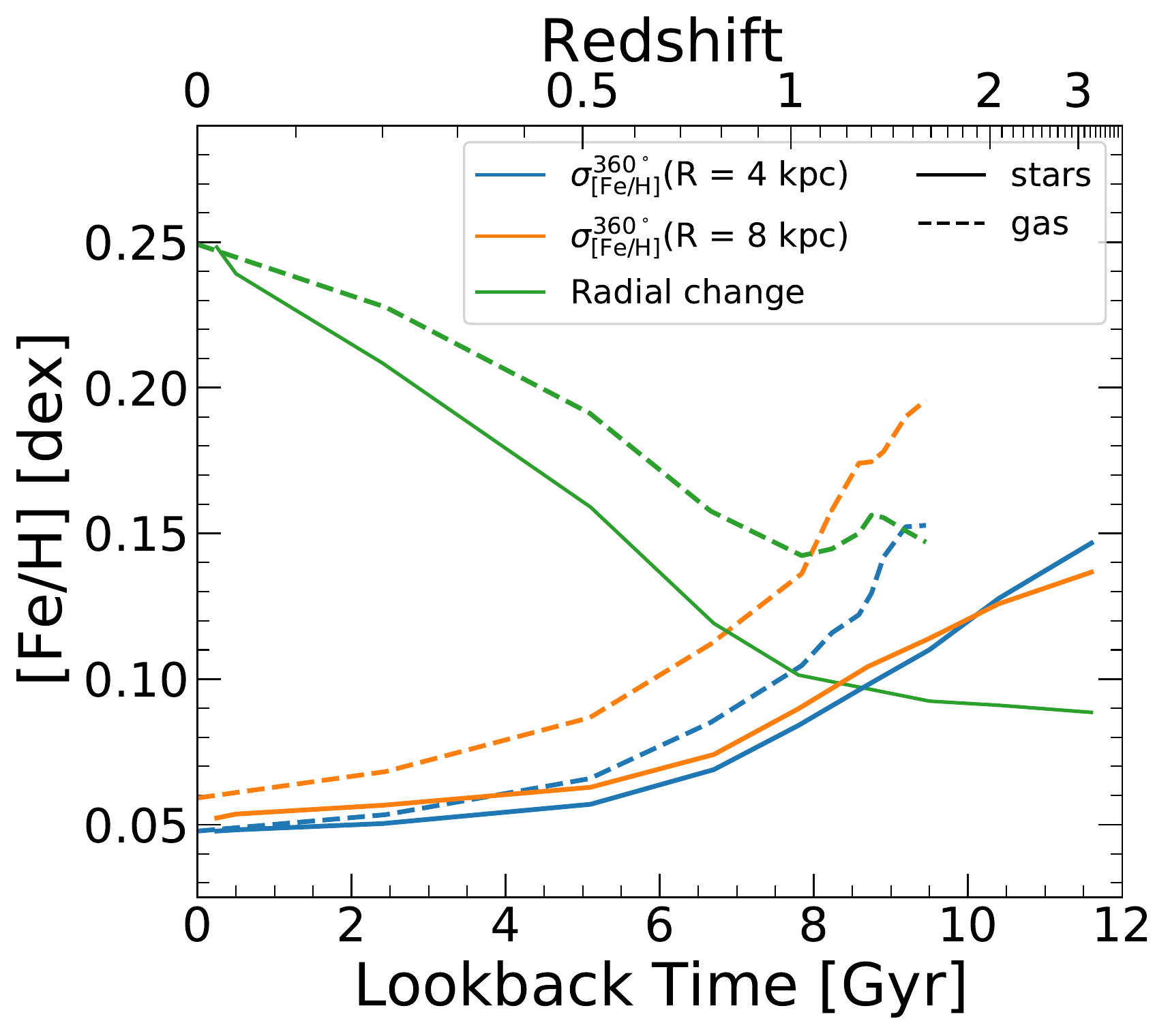}
	\vspace*{-3.5 mm}
    \caption{
    Similar to Fig.~\ref{fig:azimuthal_scatter_vs_gradient}, comparing the results for elemental abundance variations in gas versus young (age $< 500 \Myr$) stars. We generate the profiles in the same way as in Fig.~9 of \citetalias{Bellardini21} and we smooth all profiles with a Gaussian filter.  In general, both the azimuthal scatter and the radial change are smaller for newly formed stars than for all gas. Also, the transition time (see Section~\ref{subsec:azimuthal_v_radial}) is generally earlier for young stars than for all gas.
    }
    \label{fig:azimuthal_v_grad_gas_v_stars}
\end{figure}


We compare the $360^\circ$ azimuthal scatter and the strength of the radial gradient of newly formed stars (age $< 500 \Myr$) to that of all gas, as we presented in \citetalias{Bellardini21}. 
For consistency with \citetalias{Bellardini21}, instead of examining the radial change in \FeH{} from $R = 0 \kpc$ to $R^{*}_{90}$, we measure the radial gradient as a linear profile from $R = 4 \kpc$ out to $R = 12 \kpc$ and multiply by $8 \kpc$ to define a radial change in \FeH.

Fig.~\ref{fig:azimuthal_v_grad_gas_v_stars} shows the mean radial change (green) and the $360^{\circ}$ scatter at $R = 4 \kpc$ (blue) and $8 \kpc$ (orange) for young stars (solid) and gas (dashed). We find excellent agreement at $z \sim 0$, which we confirm by comparing the radial profiles of young stars and gas directly.
However, young stars show systematically lower variations, both radial and azimuthal, at all lookback times.
The largest discrepancy in the radial gradient is $\approx 0.009 \dpk$ $\approx 9.5 \Gyr$ ago. We believe this is because the stellar gradients are ill defined in this radial range for young stars at this lookback time.  As Fig.~\ref{fig:galaxy_property_evolution} shows, $R^{*}_{90}$ is only $\approx 5.5 \kpc$ on average.  The greater agreement between azimuthal scatter for gas and stars at $4 \kpc$ compared to $8 \kpc$ also supports this.
As mentioned in Section~\ref{subsec:azimuthal_scatter}, we think that clustered star formation drives smaller azimuthal scatter in young stars than in all gas.
\citetalias{Bellardini21} showed that star-forming gas in general has a smaller disk-wide scatter in abundance (median of $\sim 0.05 \dex$ at $z = 1$). This agrees with the typical difference in azimuthal scatter at large lookback times in Fig.~\ref{fig:azimuthal_v_grad_gas_v_stars}.


\section{Measurable homogeneity}
\label{appendix:measurable_homogeneity}

\begin{figure}
	\includegraphics[width = .95 \columnwidth]{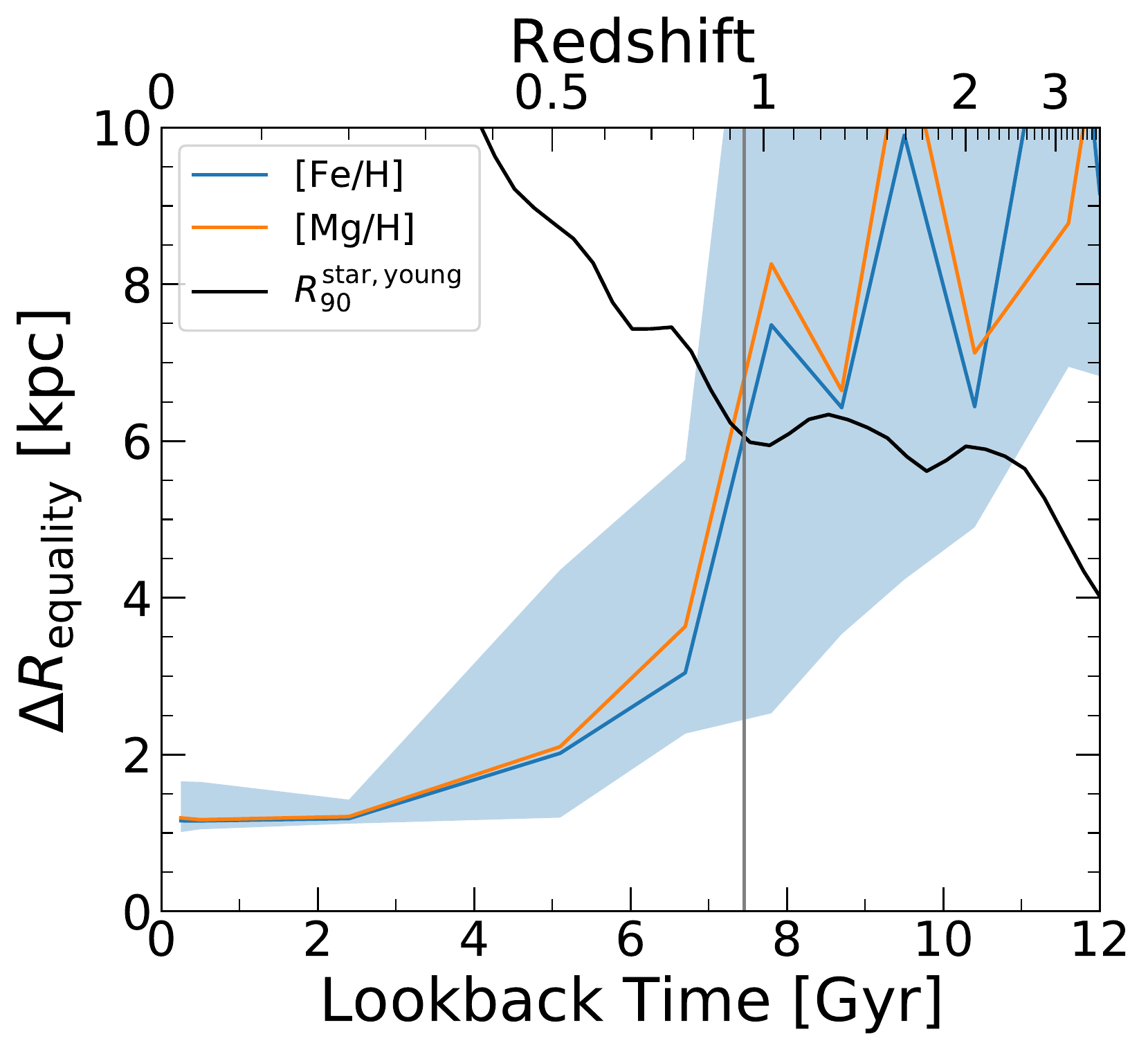}
	\vspace*{-3.5 mm}
    \caption{
    The lines show the median $\Delta R_{\rm equality}$ as a function of lookback time for \FeH{} (blue) and \MgH{} (orange) and the shaded region shows the 68th percentile for \FeH{}. For comparison, the black line shows the average $R^*_{90}$ of newly formed stars at each lookback time. We define $\Delta R_{\rm equality}$ as the ratio of the $360^{\circ}$ azimuthal scatter to the radial gradient. This ratio gives a radial scale over which azimuthal variations dominate over radial variations for \FeH{} (blue) and \MgH{} (orange). $\Delta R_{\rm equality}$ effectively sets the precision to which birth radii of stars can be measured if azimuthal scatter is neglected. For stars that formed prior to $\approx 7.5 \Gyr$ ago (right of the gray line), the azimuthal scatter dominated over radial variations across the entire galaxy. However, for stars that formed within the past $\sim 6 \Gyr$, azimuthal scatter only dominates on scales $\lesssim 3 \kpc$, less than half of $R^{*}_{90}$.
    }
    \label{fig:scatter_gradient_ratio}
\end{figure}

\begin{figure}
	\includegraphics[width = .95 \columnwidth]{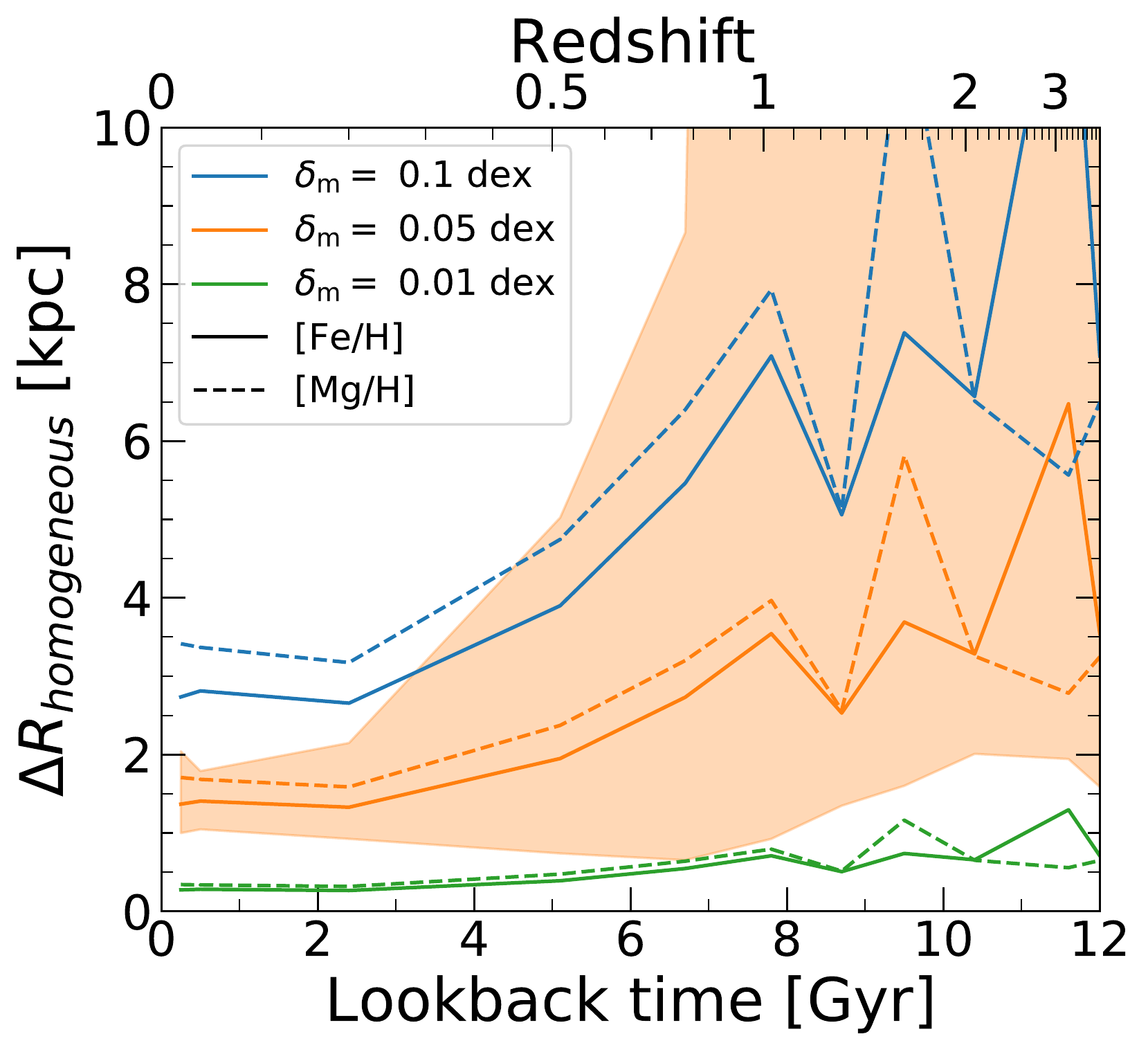}
	\vspace*{-3.5 mm}
    \caption{
    $\Delta R_{\rm homogenous}$ as a function of lookback time. $\Delta R_{\rm homogenous}$ is the ratio of assumed measurement uncertainties (taken as representative of typical observational uncertainties) to the fiducial radial gradients in our 11 simulated galaxies. The solid (dashed) line shows the median value for the simulations for \FeH{} (\MgH). The orange shaded region shows the full distribution for a fiducial scatter of $0.05 \dex$. This ratio predicts the precision to which the radial birth location of a star is definable, for a given measurement precision, assuming the radial abundance gradient is primarily responsible for setting the abundance of a star. In the simplified case of no azimuthal scatter, measurement uncertainty sets the precision of inferred stellar birth radii.  Which, for our fiducial uncertainty of $\delta_{\rm m} = 0.05 \dex$, is $\Delta R_{\rm homogeneous} \lesssim 2.7 \kpc$ for $t_{\rm lb} \lesssim 8.7 \Gyr$.
    }
    \label{fig:uncertainty_gradient_ratio}
\end{figure}

Following \citetalias{Bellardini21}, we define $\Delta R_{\rm equality}$, the ratio of the (radially averaged) $360^{\circ}$ azimuthal scatter to the overall radial gradient:
\begin{equation}
    \Delta R_{\text{equality}} = \frac{\sigma_{\text{[X/H]}}}{\Delta \text{[X/H]}_{R^*_{90}} / R^*_{90}}
\end{equation}

$\Delta R_{\rm equality}$ in effect defines the minimum radial scale over which radial variations dominate over azimuthal variations in abundance.  This sets a maximum precision that chemical tagging neglecting azimuthal scatter can place on the birth radius of a star using a given abundance measurement of the star.

Fig.~\ref{fig:scatter_gradient_ratio} shows the median $\Delta R_{\rm equality}$ as a function of lookback time. The blue line shows the median ratio for \FeH, the orange line shows the median ratio for \MgH, and the black line shows the average $R^{*}_{90}$ of newly formed stars, from Section~\ref{subsec:general_properties}. The shaded region shows the $1-\sigma$ scatter for \FeH.
$\Delta R_{\rm equality}$ in general decreases with decreasing lookback time ($\approx 1.2 \kpc$ at present day and $\approx 11.2 \kpc$ $12 \Gyr$ ago), meaning that the most precision can be placed on the birth radii of stars formed within the last $\sim 2 \Gyr$.

We can, in principal, infer the birth radii of all stars born at $t_{\rm lb} \lesssim 6 \Gyr$ to within $3 \kpc$. For stars born at $t_{\rm lb} \lesssim 500 \Myr$, the uncertainty is $\approx 1.1 \kpc$ using just measured \FeH{} or \MgH.
For stars born at $t_{\rm lb} \gtrsim 7.5 \Gyr$ (right of the gray line), $\Delta R_{\rm equality}$ was larger than the size of the galaxy. This is effectively another way to define the transition age in Fig.~\ref{fig:azimuthal_scatter_vs_gradient}.

Of course, one may be able to improve on this precision using multiple abundances at once, which we will explore in future work.

Additionally, we examine the effect of observational measurement uncertainty on the precision with which chemical tagging can indicate stellar birth radius. We present results for several measurement uncertainties $\delta_{\rm m} = 0.1$, $0.05$, and $0.01 \dex$, representative of low-, medium-, and high-resolution spectroscopic surveys, for example, GALAH \citep{Buder21}. We define $\Delta R_{\rm homogeneous}$ as the ratio of the measurement uncertainty to the radial gradient:
\begin{equation}
    \Delta R_{\text{homogeneous}} = \frac{\delta_{\text{m}}}{\Delta \text{[X/H]}_{R^*_{90}}/R^*_{90}}
\end{equation}

This ratio defines the radial scale over which the stellar disk is measurably homogeneous, assuming only a radial abundance gradient. Thus, it defines the measurement-limited precision on the birth radius of a star born in a disk dominated by a radial abundance gradient.

Fig.~\ref{fig:uncertainty_gradient_ratio} shows $\Delta R_{\rm homogeneous}$ as a function of lookback time. The solid lines show the median for \FeH, and the dashed lines show the median for \MgH. We show the full distribution for our fiducial uncertainty of $\sigma_{\rm m} = 0.05 \dex$.  For all lookback times, medium-resolution surveys give $\Delta R_{\rm homogeneous} \lesssim 4.4 \kpc$, with all stars formed at $t_{\rm lb} \lesssim 8.7 \Gyr$ having $\Delta R_{\rm homogeneous} \lesssim 2.7 \kpc$.

Fig.~\ref{fig:uncertainty_gradient_ratio} and Fig.~\ref{fig:scatter_gradient_ratio} show that $\Delta R_{\rm equality} \lesssim \Delta R_{\rm homogeneous}$ for all lookback times $\lesssim 6.7 \Gyr$ for our fiducial measurement uncertainty of $0.05 \dex$.  Thus, measurement uncertainty is the limiting factor in setting the precision of birth radii for stars born less than $\sim 5 \Gyr$ ago.  However, for stars born at $t_{\rm lb} \gtrsim 6.7 \Gyr$, $\Delta R_{\rm equality}$ was $2-3 \times$ larger, so the azimuthal scatter then was more important in setting the precision of stellar birth location.


\bsp	
\label{lastpage}
\end{document}